\documentclass[sigconf,natbib=true]{acmart}
\usepackage{balance}
\usepackage{courier}
 \usepackage{graphicx}
\usepackage{stfloats}
\usepackage{amsmath}
\usepackage{amsfonts,amssymb} 
\usepackage{multirow}
\usepackage[font={bf}]{caption}
\usepackage{booktabs}
\usepackage{threeparttable} 
\usepackage{makecell}
\usepackage{CJKutf8}
\usepackage{array,ragged2e}
\newcolumntype{P}[1]{>{\RaggedRight\arraybackslash}p{#1}}

\setcopyright{none}
\renewcommand\footnotetextcopyrightpermission[1]{}
\AtBeginDocument{%
  \providecommand\BibTeX{{%
    \normalfont B\kern-0.5em{\scshape i\kern-0.25em b}\kern-0.8em\TeX}}}

\setcopyright{acmcopyright}
\copyrightyear{2018}
\acmYear{2018}
\acmDOI{XXXXXXX.XXXXXXX}

\acmPrice{15.00}
\acmISBN{978-1-4503-XXXX-X/18/06}

\begin{document}
\begin{CJK*}{UTF8}{gbsn}
\title{Search Intenion Network for Personalized Query Auto-Completion in E-Commerce}
% \title{Search Intenion Network for Personalized Query Auto-Completion in E-commerce}

\author{Wei Bao}
\authornote{Both authors contributed equally to this research.}
\email{duandao.bw@taobao.com}
\affiliation{%
  \institution{Alibaba Group}
  \country{China}
}

\author{Mi Zhang}
\authornote{Both authors contributed equally to this research.}
\email{zishu.zm@taobao.com}
\affiliation{%
  \institution{Alibaba Group}
  \country{China}
  }

\author{Tao Zhang}
\email{guyan.zt@taobao.com}
\affiliation{%
  \institution{Alibaba Group}
  \country{China}
  }

\author{Chengfu Huo}
\email{guyan.zt@taobao.com}
\affiliation{%
  \institution{Alibaba Group}
  \country{China}
  }

% \affiliation{%

% \affiliation{%

% \affiliation{%

% \affiliation{%

% \affiliation{%

% \affiliation{%

\renewcommand{\shortauthors}{Trovato and Tobin, et al.}

% When users type in the search box character by character, QAC systems suggest personalized candidate query list for the input prefix on each new key stroke.
% Users’ search prefix represents Users’ current intention, which is meaningful for QAC modeling. However,
% This helps explicitly distinguish users’ intention transfer, by modeling users’ dynamic preferences changes over time.
% To capture user’s diverse interests from historical behaviors,
\begin{abstract}
Query Auto-Completion(QAC), as an important part of the modern search engine, plays a key role in complementing user queries and helping them refine their search intentions. Today's QAC systems in real-world scenarios face two major challenges:1)intention equivocality(IE): during the user’s typing process, the prefix often contains a combination of characters and subwords, which makes the current intention ambiguous and difficult to model.2)intention transfer (IT):previous works make personalized recommendations based on users' historical sequences, but ignore the search intention transfer.However, the current intention extracted from prefix may be contrary to the historical preferences.

In this work, we propose a neural framework called SIN (short for \textit{\textbf{S}earch \textbf{I}ntention \textbf{N}etwork}) to address these two issues.Specifically,SIN integrates different types of preferences in user behaviors and the current search intention to explicitly distinguish their real interests.
Then,Transformer-based multi-view sequence modeling is introduced to exploit diverse behavior sequences with reformulation technique. For IE problem,Transformer encoder is applied to distill local information feeded by convolutional neural network to learn equivocal intention representation.For IT problem, inspired by the intuition that the space distance between encoded vectors indicates the transfer of search intention, an interest evolution network is designed to measure users’ interest transfer.SIN has been deployed on the online search engine in 1688 website for more than 3 months.Public experimental results and long-term online A/B testing results prove that SIN is superior to other competitive mainstream models.Further studies on real-world scenarios confirm that SIN can solve and IE  and IT problems effectively and efficiently.
\end{abstract}



% \ccsdesc[500]{Information systems~Query log analysis}

% \keywords{Query Auto-Completion, Search Intention, Deep Learning}

\maketitle
%%去除页眉页脚
\pagestyle{plain}

\section{Introduction}
% 这一段 介绍QAC，讲QAC的作用或者是问题
As people's requirements for information retrieval increase, search engines apply powerful query auto-completion(QAC) services to provide a ranked list of query suggestions for users to match their intentions.When users type in the search box, QAC will immediately provide a list of recommended queries starting with the prefix, which greatly saves users' search time cost and influences the search results\cite{bar2011context}.The performance of QAC has a decisive impact on the search ranking results, affecting the users' search experience and platform revenue. Therefore, QAC system has always been the focus of academic and industrial research,such as e-commerce site and social platform.

%感觉以下内容太重复了
%  QAC services can be strengthened in the following parts:(1) Personalization has become an important part of recommendation system, which is generally modeled and represented by user history sequence. However, in the QAC system, the user's current search intention may have changed, and the QAC system needs to be able to identify the change.
% (2) The user's prefix (partial query) is usually very short and ambiguous\cite{cao2008context}. It is usually composed of characters and subwords rather than complete words and query. The QAC system needs to be able to clearly represent these combined prefixes.

% 这一段 介绍QAC的分类，GEN-QAC与CTR-QAC并引到本文的方向上。
Generally, in search engines, traditional QAC system follows a two-stage method: matching and ranking. In the matching phase, a sufficient number of candidate queries matching the prefix are recalled from the history log. In the ranking stage,the candidate historical frequency features\cite{bar2011context,jiang2014learning,whiting2014recent} and semantic features\cite{mitra2015query,jiang2018neural,wang2020efficient} are used to obtain the final list ranking order. Finally, due to the limitation of display space, several top ranked candidates will be provided to users.

Recently, inspired by the successful application of neural network(NN) in natural language processing(NLP) and recommendation system(RS) tasks, methods based on neural network are widely used in QAC systems.NN-based QAC methods can be classified into two categories: query generation QAC methods(GEN-QAC)\cite{park2017neural,fiorini2018personalized} and CTR prediction QAC methods(CTR-QAC)\cite{mitra2015query,wang2020efficient,yin2020learning}. GEN-QAC generally focuses on problems of predicting unseen queries and adopts enhanced structures of sequence-to-sequence(seq2seq) to generate queries based on the prefix.CTR-QAC tackles prefix or suffix matching problems in the matching stage, and considers semantic correlation and complicated feature interactions in the ranking stage. In e-commerce, for QAC systems, candidates have many real-world scenario restrictions and are usually limited to a prepared fixed candidate query pool. Therefore, compared with GEN-QAC, CTR-QAC is more flexible and more commonly used in  e-commerce sites.

%具体介绍CTR-QAC的文献。
% However, the information in the user's behavior sequence is diverse and dynamic
Early works of CTR-QAC systems utilize statistical features such as query frequency in historical logs for candidate ranking,but lacks semantic understanding\cite{bar2011context}.Methods based on neural network are introduced into QAC systems to solve the problems of semantic understanding and sequence modeling. In the ranking stage, convolutional latent semantic model (CLSM) measures the semantic distance between prefix and suffix, and its performance is much better than the population-based baseline ~\cite{mitra2015query}.Unnormalized language modeling is applied to model the coherence between a word and its previous sequence in candidate ranking\cite{wang2020efficient}.Recent works have begun to explore how to improve the user experience by personalizing QAC in similar ways.A transformer based multi-view multi-task framework has been used to generate and rank\cite{yin2020learning}. After obtaining user sequential behavior representations using behavior level encoder and context level encoder, a CTR prediction and a query generation model are jointly trained in a unified framework by sharing the encoder part.In short, the aforementioned works
commonly concentrate more on user recent behaviors and semantic learning, and are not capable of fully mining intention transfer to accurately estimate their current interests.
% As a result, there are two major challenges in sequential recommendation that have not
% been well-addressed so far as follows.

% However, in the existing studies, capturing interest from historical features alone can not understand users' dynamic and diverse intentions. On the other hand, incomplete prefix modeling is also difficult to describe users' immediate intentions.

%在我们本工作中，我们是怎么做的
We claim that intention understanding is the key in QAC systems. Generally, user intention includes three parts: user historical intention extracted from user historical behavior sequence, user current intention extracted from user current search prefix, and intention transfer between the above.In this work, we are committed to solving two major problems: 1) intention equivocality(IE) and 2)intention transfer (IT).
\begin{itemize}
% Intention transfer poses a challenge to today's QAC system.
\item \textbf{IE}:Understanding users' current activated core intentions is the key to IT modeling and prefix semantic representation. However,when users type in the search box character by character, QAC systems should suggest personalized candidate query list for the input prefix on each new key stroke, hence in user typing process the prefix is often very short and composed of sub-words or chars, and the intention behind prefix is not clear\cite{cao2008context}.Therefore,prefix intention understanding poses a challenge to today's QAC system.

\item \textbf{IT}:It is of great importance for QAC to be able to distinguish subtle differences between users’ evolving intentions\cite{chen2021towards}. For example, “grid-shape dress” is a specific query of “dress”,but the former represents the user's detailed search preference for “grid shape”,the contained information may be completely different from the user's historical shopping interests.However,existing studies ignore the capture of IT, which leads to QAC's inability to truly perceive users' real-time intention.
\end{itemize}
% We claim that intention understanding is the key in QAC system. Generally, user intention includes three parts: user historical intention extracted from user historical behavior sequence, user current intention extracted from user current search prefix, and intention transfer between the above. Recently, the successful application of personalization in recommendation system inspired QAC systems to focus on user sequence modeling. LSTM or transformer was used for user sequence modeling in recent studies. However, the above work lacks the representation and transfer of the user's current intention, which is a waste of prefix information.

In this paper, Search Intention Network (SIN) is proposed to address these challenges.
In e-commerce, users have a variety of interaction behavior sequences, including searched queries, clicked items,purchased items and so on. 
% Here, the first two views of historical sequential behaviors are adopted as in recent studies\cite{yin2020learning}.
For each view, the transformer based encoder is performed to capture user patterns in their historical sequences, in which the reformulation technique \cite{jiang2018rin} can better extract the user's short-term activated intention. Furthermore, a candidate to history attention mechanism is introduced to 
assign time-decaying weights to historically interacted items and queries.
% select the most useful information in historical behavior.
On the other hand, in order to address the problem of intention equivocality(IE), we utilize a convolutional neural network to extract local dependencies from character embeddings, and then use transformer to obtain prefix representation.Last but not least, for the modeling of intention transfer(IT), attention mechanism of prefix to multi-view historical sequence obtains the overall representation of the historical behavior sequence from multiple perspectives, and the cosine distance between the current intention and the historical intention illustrates the intention transfer.

%这里介绍我们的贡献。
In summary, the contributions of this work can be summarized as follows:
\begin{itemize}
\item  We point out the limitations of previous studies on users' intention equivocality and intention transfer.Further, we explain their great significance and practical value for QAC systems and declare the modeling coverage of user intention in QAC systems,in which flexible and explainable concept representations can benefit the learning of how users evolve their intentions along search behaviors for query auto-completion.
% and solve the above problems by modeling user intention from multiple perspectives.
\item  We propose the framework SIN for user intention understanding to address the above problems.More specifically,based on the accurate modeling of historical preferences and current intention, interest evolution module enable the opportunity to predict the intention transfer in adjacent search periods.SIN greatly refines the intention representation 
% granularity of the neural network 
and better captures various characteristics of user interests.
% User intention specific expression and historical interest transformation are introduced to capture the diversity and evolution of users' interests, which is enhanced by a transformer-based semantic understanding network.
\item Extensive experiments are conducted on publicly available AOL search engine
logs and 1688 online search logs. The experimental results demonstrate that SIN achieve superior performance compared with state-of-the-art methods.SIN has been deployed into commercial search engines,
% , one of world’s largest business-to-business platform,
and online A/B testing results show significant business improvement brought by SIN.
\end{itemize}

\section{Related Work}
% In this section, we will introduce related works to provide background knowledge. We first introduce  traditional methods for QAC, then sequential recommendation in query auto-completion and query suggestion, and finally char level and subword level text coding model.
\subsection{Query Auto-Completion and Suggestion}
To better meet user information requirements, search engines are equipped with sophisticated query auto-completion and query suggestion services. Early related works focused on the statistical frequency and co-occurrence. Association rules\cite{fonseca2003using} and co-occurrence information \cite{fonseca2005concept,huang2003relevant} can be mined for generating lists of related queries in query suggestion. The traditional strategy Most Popular Completion (MPC)  ranks candidate query according to its popularity in history search log\cite{mitra2015query}.Inspired by the successful and wide application of neural networks,recurrent neural network(RNN)\cite{fiorini2018personalized,wang2020efficient,00060HSMZ0G23}, convolutional neural network(CNN)\cite{mitra2015query} and language model(LM)\cite{wang2020efficient} are used in the ranking stage of query auto-completion. To better understand users' search intention, recent work has focused on modeling fine-grained representations of users' historical knowledge through neural networks.A variety of information can be applied to enhance the understanding of user intents behind the query, such as context and search session information\cite{jiang2018rin,wu2018query,YangHXH23},personalization\cite{zhong2020personalized,buaduarinzua2018using,cai2014time}, and user behaviors\cite{hofmann2014eye,mitra2014user}.In our work, besides learning the representation of user interests from historical behaviors, we shed light on the transfer and representation of user's intention, and explore how to represent the user's incomplete intention.
\subsection{Sequential Recommendation}
In contemporary search engines, such as Google and 1688, sequential recommendation can help search engines understand user preferences and make personalized recommendations. According to the survey\cite{fang2019deep,DuYZQZ0LS23}, transaction-based sequential recommendation is a general modeling method. In e-commerce, sequential recommendation extracts the user intention based on various user historical behaviors, and recommends items or services which are consistent with user preferences.Sequence modeling methods include Multi-layer perceptron (MLP)\cite{wang2015learning}, RNN\cite{hidasi2015session,feng2019deep}, CNN\cite{tuan20173d} and transformer-based models\cite{wu2020sse,ZhangSFWW0023},attention mechanism is often used to select the most important part of the historical sequence\cite{zhang2018next,zhou2018deep,KulkarniMGF23}. Different from previous work, we design a new hierarchical transformer-based encoder with reformulation technique, which performs better than traditional RNN-based models.
\subsection{Word Representation Learning}
Word representation is the key to many NLP problems\cite{chen2014fast,goldberg2017neural,peters2018deep,ZhangWLZQHJPLS23}. Traditional word representation relies on distributed hypothesis\cite{harris1954distributional} and induce representations from large unlabeled corpora using word co-occurrence statistics\cite{zhu2019systematic,LiZ23}. Due to the limitation of learning beyond distribution information and fixed vocabulary, recent works\cite{zhao2018generalizing,banar2020character,chaudhary2018adapting,LiZ23,ParkKCHYCLRYCC23} focus on subword level information extraction, including characters, character n-grams and morphemes. Our work is more related to n-gram-based character information extraction task. We study how to encode text on n-gram character level, and obtain the representation of user's search intention using CNN and Transformers.

\section{Problem Definition}
In this section, we define the objective of this paper.Assumed that current prefix entered in the search input box is $p$.Generally, a user
has $N$ kinds of sequential interaction sequences(i.e.,browsed item sequences, searched query sequences, clicked item sequences, purchased item sequences, and so on.). Among them, the $k$-th sequence $s_k$ can be recorded as {$b_1^k$, $b_2^k$, $b^k_3$,...,$b_{len\_k}^k$}and $len\_k$ is the length of the $k$-th sequence.

Based on the above information, given the candidate set set $C$ prefixed by $p$, the objective of SIN in this paper is to estimate the probability of each candidate query $q$ in $C$ being clicked by the user, so as to determine the ranking order of the user's drop-down list according to the click probability.We further define the conditional probability for $q$ as follows:
\begin{equation}\label{metric_map}
    p(q|p;s_1;s_2;...;s_N)     
\end{equation}

\section{Search Intention Network}

\subsection{Overall Architecture}
As shown in Figure 1, SIN is mainly composed of five parts, including
1)an encoder part for candidate query,which is the semantic representation of the candidate query and also used to redistribute weights on the historical behaviors in the context-level attention mechanism;
% select the most relevant part of the historical behavior sequence in the context-level attention mechanism;
2)an encoder part for prefix, which represents the user’s current search intent via a character-level view;
3)an attention-based historical intention reformulation encoder, which wraps multiple search behavior sequence into a  historical search intention vector using a context-level transformer encoder based on reformulation representations derived from transformer outputs;
4)a search intention evolution inferencer, which represents the transfer between the user’s current search intention and the past;
5)and finally a prediction layer to model multiple pooled enhanced interest signals via MLP,and then predict the probability that current candidate query being clicked.Next, we introduce each of the modules one by one.

\begin{figure*}[ht]
    \centering
    \label{fig:framework}
    \includegraphics[width=19cm]{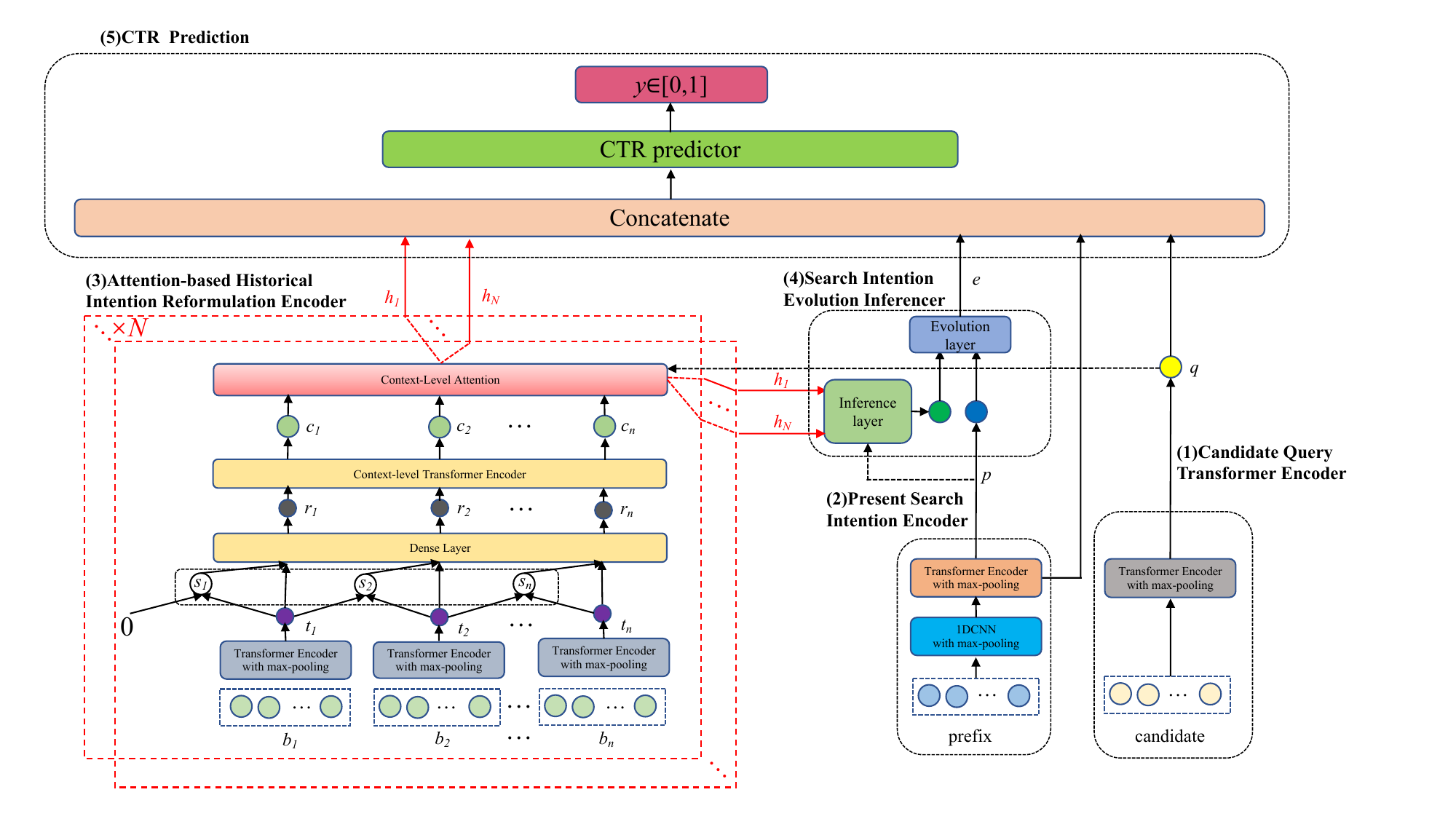}
    \caption{The schema of the proposed Search Intention Network (SIN).Followed by distilling information from the candidate query(1),entered prefix (2) and $\bf N$ kinds of user behavior sequences(3),SIN captures interest evolution between historical perference patterns and currently activated core interest(4),and finally conducts the CTR prediction task(5).}
 \end{figure*}

\subsection{Candidate Query Transformer Encoder}
In SIN, we use transformer to encode all text sequences.Traditional recurrent and convolutional models have limited receptive field and limited model capacity\cite{hochreiter2001gradient},Transformers solve the above problems with the self-attention mechanism and are widely used for processing textual input in many NLP tasks.As in \cite{vaswani2017attention},we inject positional embeddings into the input query embeddings at the bottoms of the stacked transformer layers,where we have $B$ blocks of self-attention layers and point-wise feed-forward layers, where each layer extracts features  and relationships for each time step based on the previous layer’s outputs. We omit a detailed description of the Transformer model for brevity. The detail is identical to the Transformer encoder used in the original paper (\cite{vaswani2017attention}). Finally,conventional max pooling is applied to obtain the representation vector of candidate query.

\subsection{Present Search Intention Encoder}
In real-life search scenarios, the prefix entered by the user has the characteristics of incomplete spelling, short length, and ambiguous intent. However,most existing studies
ignore users’ incompleted intentions,in this section,
% we further design a  track the evolution of users' search intention in historical interaction and current search. First, in order to get the intention of the current search
considering the incompleteness of the prefix and the difficulty of dealing with Out-Of-Vocabulary (OOV) words,we utilize a convolutional network to extract local dependencies from character embeddings and downstream apply a transformer to construct these segmented embeddings.

Specifically,for character-level convolutional neural network,let $V$ be the vocabulary of characters, $d$ be the dimensionality of character embeddings,and matrix character embeddings can be denoted as  $\textbf{E} \in \mathbb{R}^{d\times|V|}$. Suppose the prefix is made up of a sequence of character[$c_1$,...,$c_l$],where $l$ is the length of prefix.Then the character-level representation of prefix is given by the matrix $\textbf{M} \in \mathbb{R}^{d\times l}$,where the j-th column corresponds to the character embedding for $c_j$(the $c_j$-th column of E).As in [13],a narrow convolution between $\textbf{M}$ and a kernel(or filter) $\textbf{K} \in \mathbb{R}^{d\times w}$ of width $w$ followed by a nonlinearity operation is applied to to obtain a feature map $\textbf{f} \in \mathbb{R}^{l-w+1}$.Specifically, the $i$-th element of f $\textbf{f}$ is given by:
\begin{equation}\label{metric_map}
    \boldsymbol{f}[i] = \rm{tanh}(\left \langle \boldsymbol{\rm M}[*,i:i+w-1],\boldsymbol{\rm K} \right \rangle+b)
\end{equation}
where $\boldsymbol{\rm M}[*,i:i+w-1]$ is the $i$-to-$(i+w-1)$-th column of $\boldsymbol{\rm M}$ and $\left \langle \boldsymbol{\rm A},\boldsymbol{\rm B} \right \rangle$ = Tr($\boldsymbol{\rm A}\boldsymbol{\rm B}^T$) is the Frobenius inner product.
Finally, we take the \textit{max-over-time}
\begin{equation}\label{metric_map}
    \boldsymbol{y} = \max \limits_{i}\boldsymbol{f}[i]
\end{equation}
as the feature corresponding to the kernel  $\textbf{K}$.Our CharCNN uses multiple kernels of varying widths to obtain the feature vector for prefix.So if we have a total of $k$ kernels $\boldsymbol{K_1}$,$\boldsymbol{K_2}$,...,$\boldsymbol{K_k}$, then the input representation of char-level prefix can be denoted as [$\boldsymbol{y_1}$,$\boldsymbol{y_2}$,...,$\boldsymbol{y_k}$].Lastly,we feed the sequence of the processed char-level information into a second stacked Transformer layer to obtain the representation of the whole prefix $p$.
% \begin{equation}\label{metric_map}
%     \boldsymbol{p} = concatenate(y_1,y_2,...,y_h)
% \end{equation}
% where h is chosen to be 128.

\subsection{Attention-based Historical Intention Reformulation Encoder}
To obtain the user historical perference, a variety of user sequences are used for modeling to obtain the intention expression of users from different perspectives. Suppose that the user has $N$ behavior sequences.Generally,a historical sequence can be recorded as < $b_1$, $b_2$,..., $b_n$ >, which is in the text form. First, we encode each element in the sequence into a dense vector, which is built upon a transformer encoder and the traditional max pooling strategy as processed in section 4.2. Modeling query reformulation behavior has been proved to play an important role in understanding users' historical intention and intention transfer\cite{chen2021towards}.Here, we adopt simple linear operations to represent reformulations of adjacent behaviors, we define behavior-to-behavior reformulation as follows:
The reformulation $\boldsymbol {s_i}$ from $\boldsymbol{b_{i-1}}$ to $\boldsymbol{b_i}$ is defined as
\begin{equation}\label{metric_map}
    \boldsymbol{s_i} = \boldsymbol{t_i} - \boldsymbol{t_{i-1}}     
\end{equation}
Where $\boldsymbol{t_i}$ is the transformer layer output of $\boldsymbol{b_i}$. Simple linear operation helps SIN understand the changes between behaviors and facilitate the context-level encoder to learn the context information\cite{mikolov2013distributed,pennington2014glove}.Downstream, we concatenate the results of the linear operation $\boldsymbol {s_i}$ and transformer $\boldsymbol {t_i}$ at each time step and feed them to the full connection layer to get the final representation of the behavior level $\boldsymbol {r_i}$,which can be defined as:
\begin{equation}\label{metric_map}
    \boldsymbol{r_i} = \rm {ReLU}(\boldsymbol \psi_{ts}(\boldsymbol {t_i}||\boldsymbol {s_i}))   
\end{equation}
where $ \boldsymbol \psi_{ts}(\cdot)$ is the fully connected layer,ReLU($\cdot$) is the activation function for the hidden layer,and || is the concatenation operator.Then, the context-level Transformer encoder is used to learn the knowledge of context level,which models the user's historical search intention sequence. Since the importance of different search intents under the current intent is different, we introduce the attention mechanism to learn the weight of different intents in historical behavior, in which the key is the candidate query.Specifically,after passing through the context-level Transformer encoder, the output of each behavior $\boldsymbol c_i$ will first pass through the fully connected layer, and then the similarity with the key is calculated to obtain the alignment importance ${\alpha}_i$ of current $\boldsymbol c_i$. The calculation formula is as follows:
\begin{equation}\label{metric_map}
    \boldsymbol{f_i} = \rm{tanh}(\boldsymbol \psi_c(\boldsymbol{c_i})) 
\end{equation}
\begin{equation}\label{metric_map}
    \alpha_{pi} = \frac{\rm{exp}({\boldsymbol{f_i}}^T \boldsymbol{q})}{\sum_{z=1}^n \rm{exp}({\boldsymbol{f_z}}^T \boldsymbol{q})}
\end{equation}
wher $\boldsymbol \psi_c$ is the fully connected layer, and the activation function is tanh.$q$ is the output of transformer encoder for the candidate query, which is used to measure the importance of star behavior in the historical sequence. Here, a simple inner product calculation is applied to measure the similarity between vectors, and then the softmax function computes the importance of $\boldsymbol {f_i}$ in the historical sequence for the current candidate query $q$.

Finally, for each user's behavior sequence,the search intention for can be represented as the weighted sum of the context-level transformer encoder outputs $\boldsymbol h$ as follows:
\begin{equation}\label{metric_map}
    \boldsymbol{h} = \sum_{i=1}^n \alpha_{pi} \boldsymbol{c_i}
\end{equation}

\subsection{Search Intent Evolution Inferencer}
After obtaining the prefix expression $p$ and encoded user historical behavior sequences [$\boldsymbol {h_1}$,$\boldsymbol{h_2}$,...,$\boldsymbol{h_N}$], we need to model the intention transformation based on a large number of historical sequences and current intention. Firstly, in order to obtain the overall historical intention of the user, 
we implement the attention mechanism to calculate importance factors of different historical sequences on the real-time intention $p$.Then weighted sum of the historical sequence is calculated to obtain the historical expression $\boldsymbol{\widetilde{h}}$.
The attention scores between the real-time intention and the $i$-th historical sequence is then given by
\begin{equation}\label{metric_map}
    \alpha_{ei} = \frac{\rm{exp}({\boldsymbol{h_i}}^T \boldsymbol{p})}{\sum_{z=1}^N \rm{exp}({\boldsymbol{h_z}}^T \boldsymbol{p})}
\end{equation}
We then obtain a refined, high-level historical perference vector
$\boldsymbol{\widetilde{h}}$  by performing a linear combination of the original vectors as
\begin{equation}\label{metric_map}
    \boldsymbol{\widetilde{h}} = \sum_{i=1}^N \alpha_
    {ei} \boldsymbol{h_i}
\end{equation}

To further model the transfer between historical intention and current user intention, the evolution layer is introduced to measure the difference between $\boldsymbol{\widetilde{h}}$ and $\boldsymbol{p}$. Specifically, it is calculated as:
\begin{equation}\label{metric_map}
    \boldsymbol{e} = \rm {ReLU}(\boldsymbol {\widetilde{h}}-\boldsymbol{p}||\boldsymbol{\widetilde{h}}*\boldsymbol{p}||cosine(\boldsymbol{\widetilde{h}},\boldsymbol{p}))
\end{equation}
where || represents concatenation operation.Here,cosine($\boldsymbol{\widetilde{h}},\boldsymbol{p}$) is defined as:
\begin{equation}\label{metric_map}
    cosine(\boldsymbol{\widetilde{h}},\boldsymbol{p}) = \frac{\boldsymbol{\widetilde{h}}^\mathsf{T}\boldsymbol{p}}{\Vert \boldsymbol{\widetilde{h}} \Vert \Vert \boldsymbol{p} \Vert}
\end{equation}
where $\rm ||\cdot||$ is the Frobenius norm\cite{van1996matrix}.
\subsection{Prediction and optimization}
Lastly, all the vectors are concatenated together to obtain the overall representation vector for the instance.Given the dense concatenated vector, fully connected layers are used to learn the combination of features automatically. A three-layer feedforward neural network is used as the prediction function to estimate the probability of the user clicking candidate query at the next moment, and all baselines in section 5 will share this prediction function.
\begin{equation}\label{metric_map}
    p(\textbf{x}) = \boldsymbol{\rm{Predict}}(\boldsymbol {h_1}||\boldsymbol{h_2}||\boldsymbol{\cdots}||\boldsymbol{h_N}||\boldsymbol{p}||\boldsymbol{e}||\boldsymbol{q})
\end{equation}
where || represents concatenation operation,$\textbf{x}$ is the input of the network and $p(x)$ is the output of the network after the softmax layer, representing the predicted probability of sample $\textbf{x}$ being clicked.The objective function used in SIN is the negative log-likelihood function with a \textit{L2} regularization term to prevent over-fitting, which is defined as:
\begin{equation}\label{metric_map}
    \boldsymbol{L} =-\frac{1}{|Q|} \sum_{(\textbf{x},y)\in Q}(ylogp(\textbf{x})+(1-y)log(1-p(\textbf{x})))+\lambda\|\Theta\|_2
\end{equation}
$Q$ is the set of samples in the training dataset, |$Q$| is the total number of samples in the set,and $y \in \{0,1\}$ is the ground truth label.$\Theta$ denotes the set of trainable parameters and $\lambda$ controls the penalty strength.

\section{EXPERIMENTS}
In this section, we compare our proposed algorithm, Seatch Intention Network (SIN) with other state-of-the-art algorithms on two real-world datasets and online A/B testing. Experimental details include datasets, strong baseline methods, results, and discussions across several important dimensions.

\subsection{Datasets and Experimental Setup}
A large number of experiments were performed on the public data set AOL data set and a real-world dataset collected from the 1688 search log to evaluate our proposed method. 
\subsubsection{AOL search logs}
As the largest available search log, AOL search logs are widely used in related academic research on query auto completion and related studies\cite{bar2011context,jiang2016classifying,zamani2017relevance,zamani2016estimating,jaech2018personalized,wang2020efficient,mitra2015query,park2017neural}.The AOL search log contains three months of website search logs from 1 March, 2006 to 31 May, 2006. Keeping consistent with existing studies \cite{mitra2015query,wang2020efficient}, rare queries with a frequency of less than 3 and empty queries are removed.
% The sequence composed of user's last five search queries is used to describe the user's historical intention.
All the data from 1 March, 2006 to 30 April, 2006 are used as the background data, with the following two weeks as training data, and each of the following two weeks as validation and test data.This results in 14.05 million background queries, 3.50 million training queries, 1.39 million validation queries, and 1.73 million test queries.
\subsubsection{1688 Dataset}
In real industrial applications, QAC models are usually trained using real prefix-query pairs collected from online search log, in which the clicked query after the user enters the prefix is regarded as a positive sample and the rest is regarded as negative samples\cite{yin2020learning}. Due to the lack of prefix-query click behaviors in AOL query log datasets, we further conduct experiments on a large-scale query log dataset collected from 1688$\footnote{https://www.1688.com/}$,the largest B2B (Business-to-Business) e-commerce platform in China.Compared with existing AOL query log datasets, 1688 dataset has rich real prefix-to-query click behaviors.Consistent with online large-scale QAC systems, personalization and intention modeling are the key for better model performance in 1688 dataset.
% In addition, as an e-commerce scenario dataset, personalization and intention are the key for better performance, which is also consistent with large-scale search scenarios.

We collected traffic logs from the online query auto-completion system in 1688 website between July 1, 2023 and July 13, 2023,of which first 10 days of data is used for training,5,000,000 logs sampled from July 11, 2023 are used for validation, and the other logs are used for testing.
We normalize all the queries in the dataset by removing any punctuation characters and ensure rationality in e-commerce.Finally, we have 22.11 million training queries, 3.51 million validation queries,and 8.74 million test queries.

To further verify the performance of SIN in solving the problems of IE and IT, we classify 1688 dataset into two categories:1) Non-IE and IE: for prefix-query pairs in the test set, if the prefix does not contain complete product-core words or modifiers, it indicates that the search intention is incomplete and ambiguous, which is recorded as IE, otherwise it is called Non-IE.2) Non-IT and IT:
Similarly, if the mapped item category of prefix is different from the categories of historical queries (query category can be acquired from 1688 search engine), or the word segmentation results of the two belong to different categories,then such data is recorded as IT, otherwise it is called Non-IT.Finally, 2.12 million IE pairs and 3.39 million IT pairs are collected for evaluation.

\subsubsection{Implementation Details}
SIN requires a significant number of hyper-parameters to tune owing to the complex
nature of training pairs and usage of Transformers.Training details,model architecture ,text tokenization and user features details are shown in this section.

\paragraph{Training} 
We use Tensorflow$\footnote{https://www.tensorflow.org/}$ for all our experiments and conduct all experiments on a server with 32-core Intel CPU and four Nvidia GTX 1080 GPUs.The models are trained by minimizing the negative conditional log-likelihood using the Adam optimizer\cite{kingma2014adam} with an initial learning rate of 0.001 and the Noam decay schedule\cite{popel2018training}. We monitor the model loss and MMR performance every 10K training iterations on the whole validtion dataset, and training will stop whenever the monitored loss does not decrease or MMR does not improve for 5 consecutive measurements.On AOL dataset we use a batch size of 32 and on 1688 dataset we use a batch size of 128 for efficiency.

% https://github.com/microsoft/LightGBM
\paragraph{Architecture}
For the prefix CNN encoder,the filter width is chosen as 1,2,3;and the mapped filter matrices number is 50,100,100.Unless otherwise stated,we use a hidden layer
dimension of 128 in the Transformer hidden layers and the prediction
layers,in which ReLU is chosen as the activation function;except the final layer,where we use softmax to produce a probability-like distribution over class predictions. The dimension of embedding vector is 32 and 64 for prefix and complete queries, and the number of hidden neurons for Transformers is set as 128.Output layers of MLP is set as 256 × 128 × 64 × 2.Both the historical item and query sequence encoding Transformers use 6 stacked Transformer layers and 8 attention heads.The prefix and candidate Transformer encoder have 4 stacked Transformer layers and 4 attention heads.

\paragraph{Tokenization and Features}
Throughout our experiments,we use the the Chinese word segmentation service provided by Aliyun$\footnote{https://www.aliyun.com/}$ to tokenize user queries and clicked item titles.After tokenization, we truncate or use a [PAD] token to pad all sequences to 8 and 15 tokens for queries and item titles, as we found this length suffice for 97\% of queries and 99\% item titles,which provides a good trade-off in MMR performance and efficiency.
For AOL dataset,the previous 5 searched queries are acted as user historical sequences. For 1688 dataset, the previous 10 searched queries and 15 clicked item titles are acted as user historical sequence,as we found this to be more than enough(a detailed analysis is provided in Section 5.5).Historical popularity information is added to all models to enhance performance.

\subsubsection{Evaluation}
Mean Reciprocal Rank (MRR) score\cite{voorhees1999trec} is employed as the evaluation metric. Given the set of prefixes P in the test dataset, let $Y(p)$ be a ranked list of queries prefixed by $p$ determined by a ranking method. We use $rank_p$ to denote the rank of the first clicked query in $Y(p)$.Formally,
\begin{equation}\label{mmr}
    \boldsymbol{\rm MRR} = \frac{1}{{P}} \sum_{(p)\in P}\frac{1}{rank_p}
\end{equation}
Essentially, the MRR score summarizes the ranks of the first clicked
queries in the recommendation list.A larger score indicates that the clicked query is ranked higher in the candidate list.To evaluate statistical significance, we use paired student’s t-test (p<0.05) for MMR evaluation in all cases.

\subsection{Competitive Baseline Methods}
In this section, we evaluate the MRR performance of SIN against the following state-of-the-art methods.
\begin{itemize}
\item \textbf{MPC}\cite{bar2011context}:MostPopularCompletion(MPC) is a statistical model based on historical frequency. Candidate queries are ranked according to the frequency of appearance in historical logs.
\item \textbf{MCG}\cite{wang2020efficient}.Based on the existing suffix matching method \cite{mitra2015query},Maximum Context Generation(MCG) greedily match the longest suffix to generate candidates.
\item \textbf{LightGBM}\cite{ke2017lightgbm}: LightGBM is a algorithm that is widely used in industry. We represent the prefix, context and query candidate with Bag-of-Words (BoW) vectors to train a LightGBM model.
\item \textbf{CLSM}\cite{mitra2015query}: Convolutional latent semantic model(CLSM) extracts the n-gram information for the prefix and suffix after corresponding embedding representation, and computes the cosine similarity between the pooled prefix and suffix vector to obtain the ranking score of the candidate query.
\item  \textbf{ALE}: Attention-based LSTM Encoder(ALE) is applied to semantically represent user historical text sequences, prefix, and candidates.And the attention mechanism is employed to encode the search intention by reading multiple historical sequences.
\item  \textbf{ATE}: Based on ALE model,Attention-based Tansformer Encoder(ATE) utilizes Transformer instead of LSTM to represent the behavior sequence.
\item $\boldsymbol{\rm M^2A}$\cite{yin2020learning}: Multi-view Multi task framework({$\rm M^2A$}) is one of the state-of-the-art ranking models.A CTR prediction and a query generation model are jointly trained in a unified framework by sharing the encoder part.Here we only implement the discriminative part of $\rm M^2A$ without extra generation task or additional information.
\end{itemize}

\renewcommand{\arraystretch}{1.5} %控制行高  
\begin{table*}[ht]  
  \centering  
  % 6.5  8 
  \fontsize{8}{6}\selectfont  
  \begin{threeparttable}  
  \caption{Model MRR performance comparison for seen/unseen,IE/Non-IE and IT/non-IT cases on AOL dataset and 1688 dataset.The bold values denote the best results and * denotes statistically significant results through a paired t-test with p<0.05.} 
  \label{tab:performance_comparison} 
  \setlength{\tabcolsep}{4mm}{
    % \begin{tabular}{cllll}  
    \begin{tabular}{P{10mm} P{10mm} P{10mm} P{10mm} P{10mm}P{10mm} P{10mm} P{10mm} P{12mm}}
    \toprule  
    \multirow{2}{*}{\textbf{Method}}&  
    \multicolumn{2}{c}{\textbf{AOL}}&\multicolumn{6}{c}{\textbf{1688}}\cr  
    % \cmidrule(lr){2-4} \cmidrule(lr){5-7}  
    \cmidrule(lr){2-3} \cmidrule(lr){4-9}
    &\textbf{Seen}&\textbf{Unseen}&\textbf{Seen}&\textbf{Unseen}&\textbf{IE}&\textbf{Non-IE}&\textbf{IT}&\textbf{Non-IT}\cr  
    \midrule  
    \textbf{MPC}&0.4521&0.0000&0.5212&0.0000&0.1917&0.1924&0.1974&0.1977\cr   \textbf{MCG}&0.4544&0.2615&0.5235&0.4503&0.4825&0.4809&0.4811&0.4814\cr    \textbf{LightGBM}&0.4551&0.2627&0.5329&0.4862&0.5011&0.5092&0.5159&0.5182\cr  
    \cmidrule(lr){1-9}   \textbf{CLSM}&0.4557&0.2791&0.5450&0.5011&0.5136&0.5248&0.5261&0.5290\cr    
    \textbf{ALE}&$\boldsymbol{0.4559^*}$&0.2798&0.5790&0.5249&0.5391&0.5504&0.5292&0.5574\cr   
    \textbf{ATE}&$\boldsymbol{0.4559^*}$&0.2802&0.5805&0.5284&0.5405&0.5539&0.5296&0.5575\cr   
    \textbf{$M^2A$}&0.4558&0.2801&$0.5877^*$&0.5508&0.5637&$\boldsymbol{0.5799^*}$&0.5641&0.5812\cr 

    \textbf{SIN}&\textbf{0.4557}$^*$&\textbf{0.2803}$^*$&\textbf{0.5912}$^*$&\textbf{0.5652}$^*$&\textbf{0.5744}$^*$&\textbf{0.5799}$^*$&\textbf{0.5831}$^*$&\textbf{0.5860}$^*$\cr
    % \textbf{SIN}&$0.4557^*$&$\boldsymbol{0.2803^*}$&$\boldsymbol{0.5912^*}$&$\boldsymbol{0.5652^*}$&$\boldsymbol{0.5744^*}$&$\boldsymbol{0.5799^*}$&$\boldsymbol{0.5831^*}$&$\boldsymbol{0.5860^*}$\cr  
    \bottomrule  
    \end{tabular}}  
    \end{threeparttable}  
\end{table*} 

    % ALE&0.4559&0.2798&0.5790(+0.696\%)&0.5249(+0.696\%)\cr  
    % ATE&0.4559&0.2802&0.5805(+0.957\%)&0.5284(+0.696\%)\cr  
    % M^2A&0.4558&0.2801&0.5877(+2.209\%)&0.5405(+0.696\%)\cr
    % SIN&{\bf 0.4559}&{\bf 0.2804}&{\bf 0.5912(+2.817\%)}&0.5652(+0.696\%)\cr  

\subsection{Results on Offline Dataset}
Table 1 shows the results on AOL dataset and 1688 dataset.Consistent with the prior work, models are evaluated on seen and unseen queries separately according to whether the query can be matched in the background dataset.Moreover,similar experiments are conducted on real-world IE and IT datasets to verify the model performance on solving these challenging problems.
\subsubsection{Results on AOL Dataset}
For three traditional non-neural methods, MCG prioritizes the generation of candidates that share more context with the user input, and the result shows MCG is able to expand more context in user input to generate candidates with higher quality. LightGBM adds the Bag-of-Words vector and historical features of query on the generation candidate set of MCG, and the effect is further improved.

All neural ranking methods are performed on the top-ranked candidates generated by MCG.For seen queries,since the construction of samples in AOL logs does not include any real prefix-query click behaviors, the performance of personalized methods (including ALE, ATE, $\rm M^2A$,SIN) that depend on user sequence is similar to CLSM, indicating that the behavior sequence in AOL logs is not the main factor to improve performance.On the other hand,
neural networks have greatly improved in unseen queries compared with non-neural methods, indicating their strength in semantic understanding and modeling correlation between candidates and words that are not used by MCG.
\subsubsection{Results on 1688 Dataset}
Due to the lack of user real-world browsing and clicking records in AOL dataset, we further carry out extensive experiments on 1688 offline logs.Compared with AOL dataset, 1688 dataset contains a large number of real prefix-query pairs and personalization plays a key role for better performance.Looking at the ’Seen’ and ’Unseen’ column, firstly, LightGBM captures additional query semantic information based on MCG generation candidates, therefore the effect is better than MCG and MPC.Secondly,besides historical frequency, CLSM captures the semantic correlation between prefix and candidate, and performs better than these frequency-based method. Then the performance of all personalized models is better than that of non-personalized models, which proves the importance of user interest for intention modeling in 1688 dataset,this is also consistent with previous studies\cite{yin2020learning}. Finally, for unseen queries, SIN is able to model the transfer between prefix and historical behavior information, which weakens the impact of the large semantic gap between prefix and candidate query and achieves the best performance.

Looking at the ’IE’ column and ’IT’ column, we see that, the MMR performance of
SIN and all baselines significantly drops in these challenging
situations. However, SIN still significantly outperforms the competitive baseline methods in almost all cases.For the IE/Non-IE cases,an interesting observation is made for $\rm M^{2}A$ baseline, $\rm M^{2}A$ performs best in Non-IE cases ,which suggests that  $\rm M^{2}A$ is able to handle usual cases. But $\rm M^{2}A$ performs poorly on IE cases with the lack of effective prefix modeling.Instead,SIN achieves the best performance compared with these baselines in IE task.Another notable observation is further made for IT cases, where SIN struggles to capture user search intention shift, but still manages to capture interest evolution and beats the $\rm M^{2}A$ by 3.37\% and CLSM by 10.83\%.

\renewcommand{\arraystretch}{1.5} %控制行高  
\begin{table}[tp]  
  \centering  
  \fontsize{8}{6}\selectfont  
  % \begin{twoparttable}  
  \caption{Ablation study of the key designs.}  
  \label{tab:performance_comparison}  
    \begin{tabular}{cccccccccc}  
    \toprule  
    \multirow{2}{*}{\textbf{Method}}&  
    \multicolumn{4}{c}{\textbf{1688}}\cr  
    \cmidrule(lr){2-5} 
    &\textbf{IE}&\textbf{Non-IE}&\textbf{IT}&\textbf{Non-IT}\cr  
    \midrule  
    $SIN_b$&0.5386&0.5525&0.5306&0.5583\cr 
    $SIN_b+H$&0.5725&0.5764&0.5749&0.5796\cr 
    $SIN_b+H+SE$&0.5730&0.5769&{\bf 0.5831}&0.5843\cr 
    $SIN_b+H+PI$&{\bf 0.5745}&0.5792&0.5747&0.5827\cr 
    $SIN$ &0.5744&{\bf 0.5799}&{\bf 0.5831}&{\bf 0.5860}\cr  
    % $\boldsymbol{$\rm$ $SIN_b$}$&0.5386&0.5525&0.5306&0.5583\cr  
    % \boldsymbol{\rm SIN_b+H}&0.5725&0.5764&0.5749&0.5796\cr  
    % \boldsymbol{\rm SIN_b+H+SE}&0.5730&0.5769&{\bf 0.5831}&0.5843\cr  
    % \boldsymbol{\rm SIN_b+H+PI}&{\bf 0.5745}&0.5792&0.5747&0.5827\cr 
    % \boldsymbol{\rm SIN}&0.5744&{\bf 0.5799}&{\bf 0.5831}&{\bf 0.5860}\cr  
    \bottomrule  
    \end{tabular}  
    % \end{twoparttable}  
\end{table} 
\subsection{Ablation Study}
\subsubsection{Model Design}

To further evaluate the impact of different model components in SIN on solving IE and IT problems, the following modules are separated and studied separately. 1)attention-based historical intention reformulation encoder - \textbf{H}; 2)Search Intention  Evolution Inferencer -\textbf{SE}; 3)Present Search Intent encoder - \textbf{PI} and basic transformer-based model -$\rm \bf SIN_b$ that does not contain the above components.Detailed MMR results for solving IE and IT problems by different components on the 1688 dataset are summarized in Table 2.

Focusing on the detail of Table 2,first, adding historical encoder \textbf{H} to $\rm \bf SIN_b$ can not solve the above problems, but the effect is better than the baseline model, which also proves that reformulation technology is helpful to the modeling of user historical behavior sequence.Secondly, Based on $\rm \bf SIN_b$ structure, after adding \textbf{SE} and \textbf{PI} respectively, the original network is endowed with the ability to predict and understand the current intention, and the results show their effectiveness. In summary, we conclude that the SIN model we proposed can more
effectively alleviate the above issues compared with original network $\rm \bf SIN_b$.

\subsubsection{Universality Study for Key Components}
%% 在CLSM、M2A上分别增加组件，观察MMR。Ablation study of the key designs
In this section, we further show the universality of key components(see section 5.4.1) by demonstrating the metrics of utilizing  them for enhancing traditional competitive baseline models. To this end, we perform an offline evaluation of CLSM,ATE and $\rm M^{2}A$ with and without the above components.Specifically,we compare the performance of CLSM,ATE and $\rm M^{2}A$ when adding \textbf{H}, \textbf{SE} and \textbf{PI} modules against original models, and the results on 1688 dataset are shown in Figure 2.

\begin{figure}
    \centering
    \label{fig:framework}
    \includegraphics[width=0.48\textwidth]{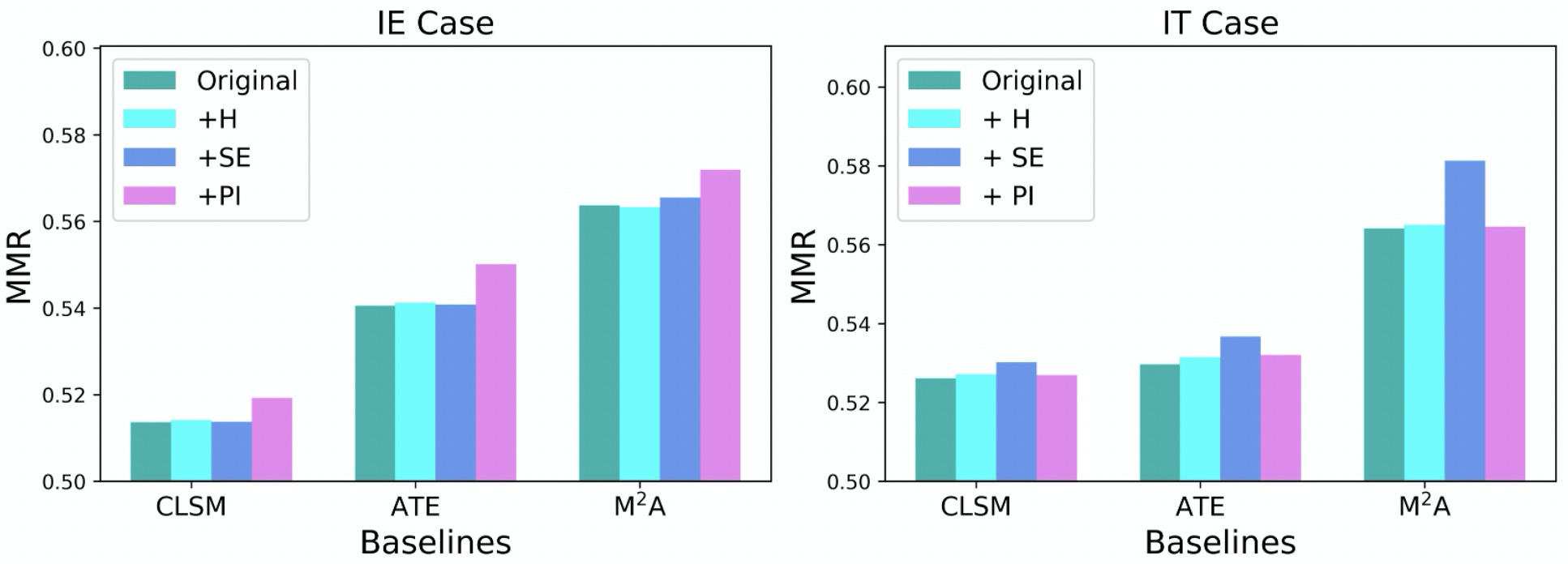}
    \caption{Performance comparison of the baseline method
adding different key designs.}
 \end{figure}

The first observation is that the performance of \textbf{H}-added models are similar with original models for both IE and IT problems, which shows that only extracting historical features cannot address these challenging problems effectively.The second observation is that adding \textbf{SE} and \textbf{PI} modules on the baselines can bring benefits to all existing methods for solving IE and IT problems,respectively. It shows that our key components will significantly reduce the difficulty of modeling user intentions and obtain better performance.As we would expect, \textbf{PI}-added models performs best in IE cases and \textbf{SE}-added models performs best in IT cases,which demonstrates the effectiveness and universality of SIN components in improving downstream query recommendation services for IE and IT cases in e-commerce.

% Finally, We demonstrate the importance and universality of SIN components in improving downstream query recommendation services for IE and IT cases in e-commerce.

\subsection{Hyper-parameter Analysis}
In this section,we change the hyper-parameters in SIN to explore the impact of hyperparametric changes on MMR performance.
\subsubsection{Length of user historical sequences}
\begin{figure}
    \centering
    \label{fig:framework}
    \includegraphics[width=0.48\textwidth]{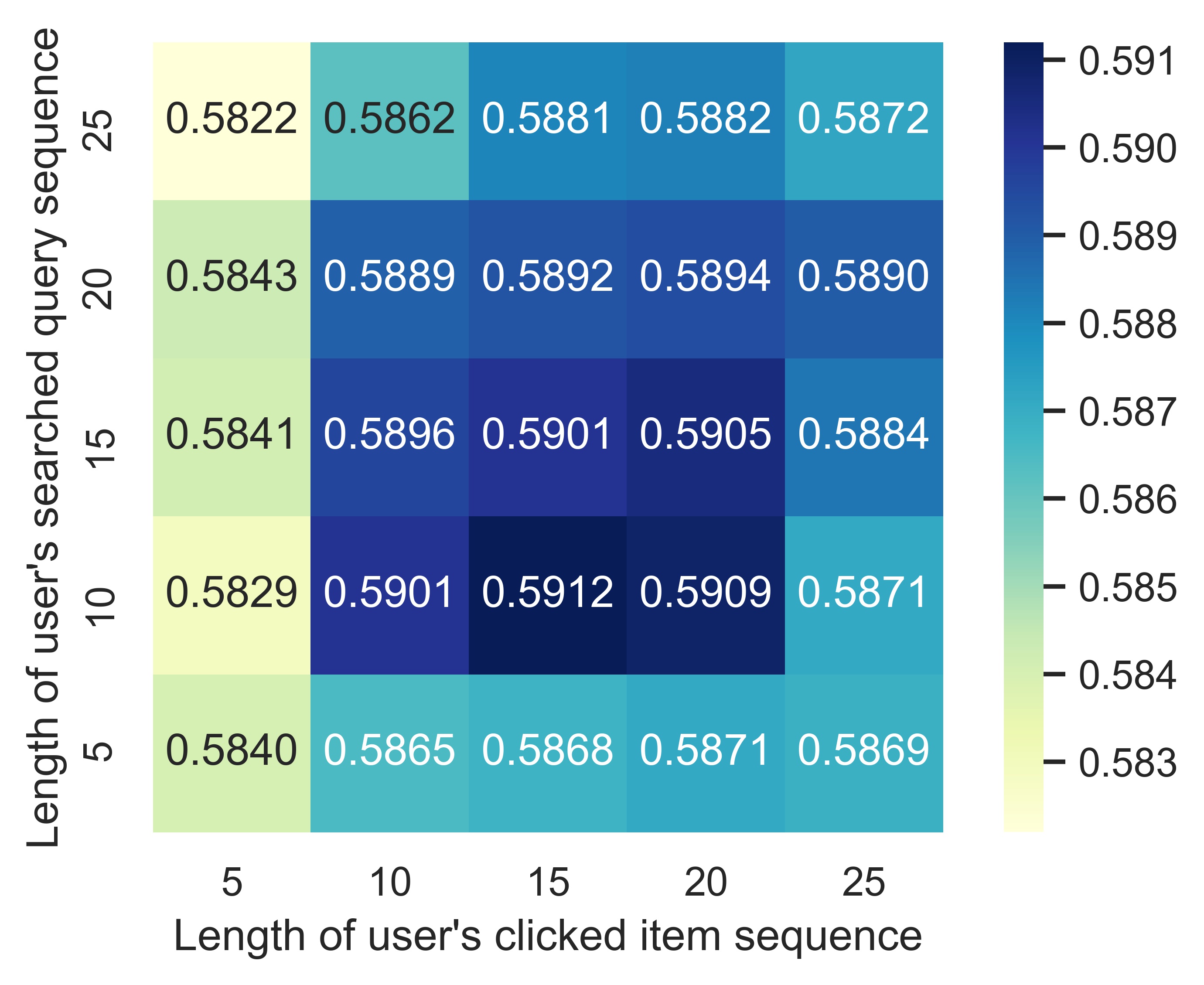}
    \caption{SIN performance by history lengths.}
 \end{figure}
%% heatmap
Diverse behavior sequences express users' rich historical interests and are an indispensable part of sequence recommendation in industry.However, too long behavior sequences may bring  additional noise signals and mislead the modeling of real interests. Thus,we study the performance of SIN on 1688 offline logs under different behavior sequence lengths to see the impact of the history length on model performances, as shown in Figure 3. Due to an extensive set of ablations of SIN fitted in the experiment, and to reduce cost and resource usage, only seen 1688 dataset are used for evaluation throughout the experiment.

%%（1）用户足迹为5时，表现一直不好。(2)过长过短都不好(3)用户搜索词稀疏。导致过长效果不好。 --- (1)(3)合并了
Firstly, compared with the query sequence, the clicked item sequence is more important for MMR performance. When the length of the user click sequence is fixed to a small value (e.g. 5), SIN has a bad performance although the length of the query sequence is changing.Because in e-commerce, the user's search behavior is sparse than the click behavior, thus the intention is not as accurate as the click behavior.
In addition, we can see retrieving either too short or too long sequences hurts the query prediction. When too few user behavior elements are retrieved, the personalized effect is limited and SIN cannot accurately predict user interest.On the other hand, when too many user interaction behaviors are retrieved, noise may be introduced and mislead the intention modeling.
\subsubsection{Embedding size of prefix}
\begin{figure}
    \centering
    \label{fig:framework}
    \includegraphics[width=0.48\textwidth]{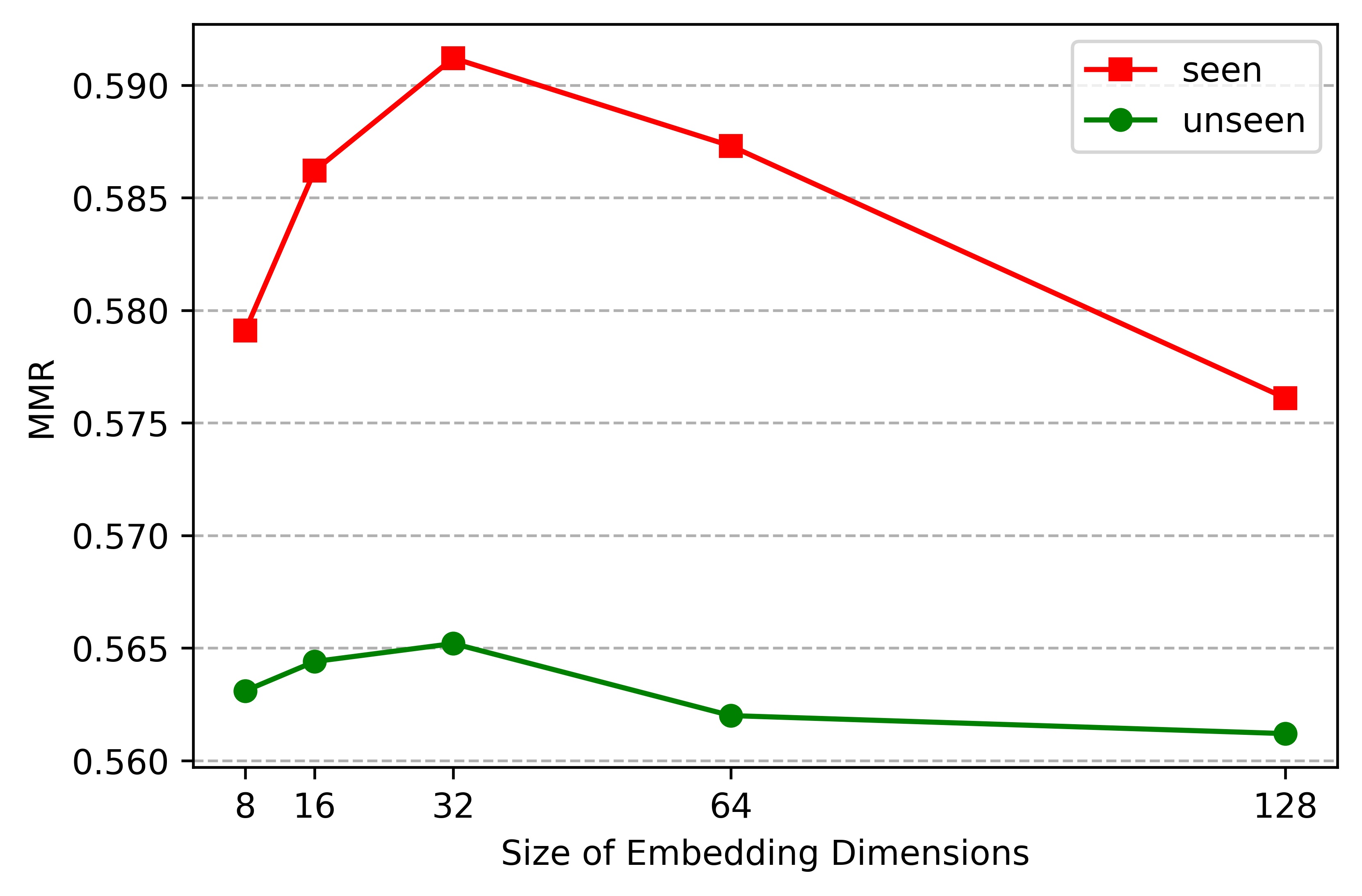}
    \caption{The MRR performance of SIN over different numbers of embedding dimensions.}
 \end{figure}
%% zhe-line
We further inspect the relationships between MMR performance and the nubmer of prefix embedding dimensions on 1688 dataset.Specifically,MMR results are recorded when prefix embedding size changes and other hyper-parameters remain unchanged.
Results are shown in Figure 4,MMR improves when the embedding dimension increases from a small value with a peak at the dimension number of 32, indicating that the information of prefix is continuously enhanced in deep semantics. However, when the embedding size is further increased,SIN becomes overfitted due to the incompleteness and the short length of the prefix. On the other hand, the effect of SIN on unseen datasets is not as sensitive to the embedding dimension as seen datasets,suggesting that prefix can only provide a limited understanding for unseen queries. Therefore,although the embedding dimension is changing, no extra deep information can be provided to extract the current search intention from the prefix.

\renewcommand{\arraystretch}{1.5} %控制行高
\begin{table*}[tp]
	\centering
	\fontsize{9}{6}\selectfont
	\begin{threeparttable}
		\caption{\textbf{Top query candidates from MPC, $M^2A$ and SIN for for the same prefix and history behavior. For case IE, the user has a historical shopping tendency to buy light bulbs, but currently only enters "L", and the prefix received by the QAC service is ambiguous.SIN models prefix representation and user real-time core interests, and can recommend the most relevant candidate queries.For case IT,the recent user sequence is all about men’s sports related items, but the user’s current prefix is 男士休闲(men’s casual), which is the shift of existing sports interest.Only 3 top-ranked queries are shown for brevity,and underlined values denote the golden results.}}
		\label{tab:performance_comparison}
		\setlength{\tabcolsep}{4mm}{
% 		\begin{tabular}{p{20pt}p{15pt}p{100pt}p{100pt}p{100pt}}
		\begin{tabular}{P{6mm} P{6mm} P{38mm} P{38mm} P{38mm}}
			\toprule
            % \makecell[c]{\textbf{Case}} & \makecell[c]{\textbf{Rank}} & \makecell[c]{\textbf{MPC}} & \makecell[c]{$\boldsymbol{\rm M^2A}$} & \makecell[c]{\textbf{SIN}} \cr
            \multirow{1}{*}{\textbf{Case}} & \multirow{1}{*}{\textbf{Rank}} &
            \makecell[c]{\multirow{1}{*}{\textbf{MPC}}} & \makecell[c]{\multirow{1}{*}{$M^2A$}} &
            \makecell[c]{\multirow{1}{*}{\textbf{SIN}}} \cr
            % &All&Seen&Unseen&All&Seen&Unseen\cr  
			\midrule
			&\makecell[c]{\textbf{1}}&LuLu连衣裙(LuLu dress)&LED灯具 家用(LED lamps household)&\underline{LED灯泡(LED bulb)}\cr
			\makecell[c]{\textbf{IE}}&\makecell[c]{\textbf{2}}&lv帽子(LV hat)&LED灯具(LED lamps)&LED 灯具(LED lamps)\cr
			&\makecell[c]{\textbf{3}}&LED灯具(LED lamps)&\underline{LED灯泡 (LED bulb)}&LED灯具 家用 (LED lamps household)\cr
			\hline
			&\makecell[c]{\textbf{1}}&男士休闲卫衣 时尚(Men's casual clothes fashion)&男士休闲跑鞋 运动(Men's casual running shoes sports)&\underline {男士休闲鞋(Men's casual shoes)}\cr
			\makecell[c]{\textbf{IT}}&\makecell[c]{\textbf{2}}&男士休闲套装(Men's casual suit)&男士休闲运动背包(Men's casual sports backpack)&男士休闲跑鞋 运动(Men's casual running shoes sports)\cr
			&\makecell[c]{\textbf{3}}&男士休闲裤(Men's casual pants)&\underline{男士休闲鞋(Men's casual shoes)}&男士休闲运动鞋(Men's casual sports shoes)\cr
			\bottomrule
		\end{tabular}}
	\end{threeparttable}
\end{table*}

\subsection{Case Study}
% \subsubsection{Case1:User Search Intention Evolution}
% Traditional QAC services focus on providing popular queries, with the development of QAC services, modern QAC services recommend personalized candidate querys based on the user’s recent behavior sequence
Real-world QAC systems aim to provide users with personalized and accurate queries to facilitate user search experience.However, users have a variety of search habits in e-commerce scenarios,modern personalized QAC systems cannot analyze the user's fuzzy intention, and tend to recommend queries related to products that the user has clicked before, even if the user's intention has changed significantly.As presented in Table 3,
1) For case IE, the QAC system needs to provide a recommendation list when the user only inputs “L”.However,previous personalized models are difficult to give accurate results when the user's input words are equivocal.SIN captures the relationship between prefix and historical behavior sequence, and has achieved extraordinary results compared with traditional methods.
2) For case IT, when the user enters 男士休闲(men's casual) after clicking (or purchasing) men's sports goods, the current immediate intention is contrary to the “sports” intention of historical shopping. MPC recommends the hottest query, and $\rm M^2A$ recommends the user's historical preference query, but ignores the current intention transfer. $\rm SIN$ can model the user's interest shift and recommend the query that meets the “casual” preference.
Therefore, we conclude that the SIN model can more efficiently model users’ real interests and  alleviate IE/IT problems effectively.
\subsection{Online Serving \& A/B testing}
A careful A/B testing was performed on the search display system of 1688 website. From July 2023 to September 2023, SIN improved CTR by 12.9\% compared with the traditional baseline model. In addition, SIN increased search unique visitors(UV) by 17.5\%. This shows the effectiveness and practicability of the proposed method. Now SIN has been deployed in 1688 search engine and serves the main traffic everyday.

In commercial search engines, it is very important for QAC system to quickly respond to user requests. However,heavy neural models are facing the pressure of difficult online deployment. The adoption of the following technologies plays an important role in the online deployment and helps SIN achieve service response within 10ms.1)\textit{User Sequence Filtering}:SIN models a variety of user behavior sequences to achieve accurate personalization. To reduce calculate consumption of 
SIN for a variety of historical sequences, we filter the user sequence and only retain the items and queries related to the current search intention. In this way,sequence  modeling consumption is saved and the similar model performance can be obtained. 2) \textit{Request batching}: adjacent service requests are merged into one batch for online inference to ensure the best utilization of GPU computing resources. Online serving of SIN benefits a lot from the above methods.

\section{CONCLUSIONS}
In this paper, we propose a novel QAC framework,namely Search Intention Network (SIN), to solve the key problems of contemporary QAC systems\,-\,intention equivocality(IE) and intention transfer(IT).
To ensure flexible personalization and remarkable effectiveness, we design a present search intention model to representate current interest state with more supervision and an reformulation encoder to encode a variety of behavior sequences.To bridge the existing gap between current intention and the history, we utilize an intention evolution inferencer to learn users' intention shift.
% Specifically, we design a present search intention encoder to representate current interest state with more supervision,where CNN is used to capture sub-word relationships and transformer is introduced for modeling.After utilizing the reformulation encoder to encode a variety of behavior sequences, an intention evolution inferencer describes the user's intention shift, which plays a key role in the design of the whole SIN framework.
The performance on offline benchmark dataset and long-term online A/B testing have proved the advantages of SIN in intention understanding.SIN has been deployed in the search system of 1688 website and undertakes the responsibility of providing users with more accurate queries to refine users' search needs.

\bibliographystyle{ACM-Reference-Format}
\balance
\bibliography{ref}


\begin{thebibliography}{58}

%%% the Web address.

\ifx \showCODEN    \undefined \def \showCODEN     #1{\unskip}     \fi
\ifx \showDOI      \undefined \def \showDOI       #1{#1}\fi
\ifx \showISBNx    \undefined \def \showISBNx     #1{\unskip}     \fi
\ifx \showISBNxiii \undefined \def \showISBNxiii  #1{\unskip}     \fi
\ifx \showISSN     \undefined \def \showISSN      #1{\unskip}     \fi
\ifx \showLCCN     \undefined \def \showLCCN      #1{\unskip}     \fi
\ifx \shownote     \undefined \def \shownote      #1{#1}          \fi
\ifx \showarticletitle \undefined \def \showarticletitle #1{#1}   \fi
\ifx \showURL      \undefined \def \showURL       {\relax}        \fi
% invisible to TeX
\providecommand\bibfield[2]{#2}
\providecommand\bibinfo[2]{#2}
\providecommand\natexlab[1]{#1}
\providecommand\showeprint[2][]{arXiv:#2}

\bibitem[B{\u{a}}d{\u{a}}r{\^\i}nz{\u{a}} et~al\mbox{.}(2018)]%
        {buaduarinzua2018using}
\bibfield{author}{\bibinfo{person}{Ioan B{\u{a}}d{\u{a}}r{\^\i}nz{\u{a}}},
  \bibinfo{person}{Adrian Sterca}, {and} \bibinfo{person}{Florian~Mircea
  Boian}.} \bibinfo{year}{2018}\natexlab{}.
\newblock \showarticletitle{Using the user’s recent browsing history for
  personalized query suggestions}. In \bibinfo{booktitle}{\emph{2018 26th
  International Conference on Software, Telecommunications and Computer
  Networks (SoftCOM)}}. IEEE, \bibinfo{pages}{1--6}.
\newblock


\bibitem[Banar et~al\mbox{.}(2020)]%
        {banar2020character}
\bibfield{author}{\bibinfo{person}{Nikolay Banar}, \bibinfo{person}{Walter
  Daelemans}, {and} \bibinfo{person}{Mike Kestemont}.}
  \bibinfo{year}{2020}\natexlab{}.
\newblock \showarticletitle{Character-level transformer-based neural machine
  translation}. In \bibinfo{booktitle}{\emph{Proceedings of the 4th
  International Conference on Natural Language Processing and Information
  Retrieval}}. \bibinfo{pages}{149--156}.
\newblock


\bibitem[Bar-Yossef and Kraus(2011)]%
        {bar2011context}
\bibfield{author}{\bibinfo{person}{Ziv Bar-Yossef} {and} \bibinfo{person}{Naama
  Kraus}.} \bibinfo{year}{2011}\natexlab{}.
\newblock \showarticletitle{Context-sensitive query auto-completion}. In
  \bibinfo{booktitle}{\emph{Proceedings of the 20th international conference on
  World wide web}}. \bibinfo{pages}{107--116}.
\newblock


\bibitem[Cai et~al\mbox{.}(2014)]%
        {cai2014time}
\bibfield{author}{\bibinfo{person}{Fei Cai}, \bibinfo{person}{Shangsong Liang},
  {and} \bibinfo{person}{Maarten De~Rijke}.} \bibinfo{year}{2014}\natexlab{}.
\newblock \showarticletitle{Time-sensitive personalized query auto-completion}.
  In \bibinfo{booktitle}{\emph{Proceedings of the 23rd ACM international
  conference on conference on information and knowledge management}}.
  \bibinfo{pages}{1599--1608}.
\newblock


\bibitem[Cao et~al\mbox{.}(2008)]%
        {cao2008context}
\bibfield{author}{\bibinfo{person}{Huanhuan Cao}, \bibinfo{person}{Daxin
  Jiang}, \bibinfo{person}{Jian Pei}, \bibinfo{person}{Qi He},
  \bibinfo{person}{Zhen Liao}, \bibinfo{person}{Enhong Chen}, {and}
  \bibinfo{person}{Hang Li}.} \bibinfo{year}{2008}\natexlab{}.
\newblock \showarticletitle{Context-aware query suggestion by mining
  click-through and session data}. In \bibinfo{booktitle}{\emph{Proceedings of
  the 14th ACM SIGKDD international conference on Knowledge discovery and data
  mining}}. \bibinfo{pages}{875--883}.
\newblock


\bibitem[Chaudhary et~al\mbox{.}(2018)]%
        {chaudhary2018adapting}
\bibfield{author}{\bibinfo{person}{Aditi Chaudhary}, \bibinfo{person}{Chunting
  Zhou}, \bibinfo{person}{Lori Levin}, \bibinfo{person}{Graham Neubig},
  \bibinfo{person}{David~R Mortensen}, {and} \bibinfo{person}{Jaime~G
  Carbonell}.} \bibinfo{year}{2018}\natexlab{}.
\newblock \showarticletitle{Adapting word embeddings to new languages with
  morphological and phonological subword representations}.
\newblock \bibinfo{journal}{\emph{arXiv preprint arXiv:1808.09500}}
  (\bibinfo{year}{2018}).
\newblock


\bibitem[Chen and Manning(2014)]%
        {chen2014fast}
\bibfield{author}{\bibinfo{person}{Danqi Chen} {and}
  \bibinfo{person}{Christopher~D Manning}.} \bibinfo{year}{2014}\natexlab{}.
\newblock \showarticletitle{A fast and accurate dependency parser using neural
  networks}. In \bibinfo{booktitle}{\emph{Proceedings of the 2014 conference on
  empirical methods in natural language processing (EMNLP)}}.
  \bibinfo{pages}{740--750}.
\newblock


\bibitem[Chen et~al\mbox{.}(2021)]%
        {chen2021towards}
\bibfield{author}{\bibinfo{person}{Jia Chen}, \bibinfo{person}{Jiaxin Mao},
  \bibinfo{person}{Yiqun Liu}, \bibinfo{person}{Fan Zhang},
  \bibinfo{person}{Min Zhang}, {and} \bibinfo{person}{Shaoping Ma}.}
  \bibinfo{year}{2021}\natexlab{}.
\newblock \showarticletitle{Towards a Better Understanding of Query
  Reformulation Behavior in Web Search}.
\newblock  (\bibinfo{year}{2021}).
\newblock


\bibitem[Du et~al\mbox{.}(2023)]%
        {DuYZQZ0LS23}
\bibfield{author}{\bibinfo{person}{Xinyu Du}, \bibinfo{person}{Huanhuan Yuan},
  \bibinfo{person}{Pengpeng Zhao}, \bibinfo{person}{Jianfeng Qu},
  \bibinfo{person}{Fuzhen Zhuang}, \bibinfo{person}{Guanfeng Liu},
  \bibinfo{person}{Yanchi Liu}, {and} \bibinfo{person}{Victor~S. Sheng}.}
  \bibinfo{year}{2023}\natexlab{}.
\newblock \showarticletitle{Frequency Enhanced Hybrid Attention Network for
  Sequential Recommendation}. In \bibinfo{booktitle}{\emph{Proceedings of the
  46th International {ACM} {SIGIR} Conference on Research and Development in
  Information Retrieval, {SIGIR} 2023, Taipei, Taiwan, July 23-27, 2023}}.
  \bibinfo{pages}{78--88}.
\newblock
\urldef\tempurl%
\url{https://doi.org/10.1145/3539618.3591689}
\showDOI{\tempurl}


\bibitem[Fang et~al\mbox{.}(2019)]%
        {fang2019deep}
\bibfield{author}{\bibinfo{person}{Hui Fang}, \bibinfo{person}{Guibing Guo},
  \bibinfo{person}{Danning Zhang}, {and} \bibinfo{person}{Yiheng Shu}.}
  \bibinfo{year}{2019}\natexlab{}.
\newblock \showarticletitle{Deep learning-based sequential recommender systems:
  Concepts, algorithms, and evaluations}. In
  \bibinfo{booktitle}{\emph{International Conference on Web Engineering}}.
  Springer, \bibinfo{pages}{574--577}.
\newblock


\bibitem[Feng et~al\mbox{.}(2019)]%
        {feng2019deep}
\bibfield{author}{\bibinfo{person}{Yufei Feng}, \bibinfo{person}{Fuyu Lv},
  \bibinfo{person}{Weichen Shen}, \bibinfo{person}{Menghan Wang},
  \bibinfo{person}{Fei Sun}, \bibinfo{person}{Yu Zhu}, {and}
  \bibinfo{person}{Keping Yang}.} \bibinfo{year}{2019}\natexlab{}.
\newblock \showarticletitle{Deep session interest network for click-through
  rate prediction}.
\newblock \bibinfo{journal}{\emph{arXiv preprint arXiv:1905.06482}}
  (\bibinfo{year}{2019}).
\newblock


\bibitem[Fiorini and Lu(2018)]%
        {fiorini2018personalized}
\bibfield{author}{\bibinfo{person}{Nicolas Fiorini} {and}
  \bibinfo{person}{Zhiyong Lu}.} \bibinfo{year}{2018}\natexlab{}.
\newblock \showarticletitle{Personalized neural language models for real-world
  query auto completion}.
\newblock \bibinfo{journal}{\emph{arXiv preprint arXiv:1804.06439}}
  (\bibinfo{year}{2018}).
\newblock


\bibitem[Fonseca et~al\mbox{.}(2005)]%
        {fonseca2005concept}
\bibfield{author}{\bibinfo{person}{Bruno~M Fonseca}, \bibinfo{person}{Paulo
  Golgher}, \bibinfo{person}{Bruno P{\^o}ssas}, \bibinfo{person}{Berthier
  Ribeiro-Neto}, {and} \bibinfo{person}{Nivio Ziviani}.}
  \bibinfo{year}{2005}\natexlab{}.
\newblock \showarticletitle{Concept-based interactive query expansion}. In
  \bibinfo{booktitle}{\emph{Proceedings of the 14th ACM international
  conference on Information and knowledge management}}.
  \bibinfo{pages}{696--703}.
\newblock


\bibitem[Fonseca et~al\mbox{.}(2003)]%
        {fonseca2003using}
\bibfield{author}{\bibinfo{person}{Bruno~M Fonseca},
  \bibinfo{person}{Paulo~Braz Golgher}, \bibinfo{person}{Edleno~Silva de
  Moura}, {and} \bibinfo{person}{Nivio Ziviani}.}
  \bibinfo{year}{2003}\natexlab{}.
\newblock \showarticletitle{Using association rules to discover search engines
  related queries}. In \bibinfo{booktitle}{\emph{Proceedings of the IEEE/LEOS
  3rd International Conference on Numerical Simulation of Semiconductor
  Optoelectronic Devices (IEEE Cat. No. 03EX726)}}. IEEE,
  \bibinfo{pages}{66--71}.
\newblock


\bibitem[Goldberg(2017)]%
        {goldberg2017neural}
\bibfield{author}{\bibinfo{person}{Yoav Goldberg}.}
  \bibinfo{year}{2017}\natexlab{}.
\newblock \showarticletitle{Neural network methods for natural language
  processing}.
\newblock \bibinfo{journal}{\emph{Synthesis lectures on human language
  technologies}} \bibinfo{volume}{10}, \bibinfo{number}{1}
  (\bibinfo{year}{2017}), \bibinfo{pages}{1--309}.
\newblock


\bibitem[Harris(1954)]%
        {harris1954distributional}
\bibfield{author}{\bibinfo{person}{Zellig~S Harris}.}
  \bibinfo{year}{1954}\natexlab{}.
\newblock \showarticletitle{Distributional structure}.
\newblock \bibinfo{journal}{\emph{Word}} \bibinfo{volume}{10},
  \bibinfo{number}{2-3} (\bibinfo{year}{1954}), \bibinfo{pages}{146--162}.
\newblock


\bibitem[Hidasi et~al\mbox{.}(2015)]%
        {hidasi2015session}
\bibfield{author}{\bibinfo{person}{Bal{\'a}zs Hidasi},
  \bibinfo{person}{Alexandros Karatzoglou}, \bibinfo{person}{Linas Baltrunas},
  {and} \bibinfo{person}{Domonkos Tikk}.} \bibinfo{year}{2015}\natexlab{}.
\newblock \showarticletitle{Session-based recommendations with recurrent neural
  networks}.
\newblock \bibinfo{journal}{\emph{arXiv preprint arXiv:1511.06939}}
  (\bibinfo{year}{2015}).
\newblock


\bibitem[Hochreiter et~al\mbox{.}(2001)]%
        {hochreiter2001gradient}
\bibfield{author}{\bibinfo{person}{Sepp Hochreiter}, \bibinfo{person}{Yoshua
  Bengio}, \bibinfo{person}{Paolo Frasconi}, \bibinfo{person}{J{\"u}rgen
  Schmidhuber}, {et~al\mbox{.}}} \bibinfo{year}{2001}\natexlab{}.
\newblock \bibinfo{title}{Gradient flow in recurrent nets: the difficulty of
  learning long-term dependencies}.
\newblock
\newblock


\bibitem[Hofmann et~al\mbox{.}(2014)]%
        {hofmann2014eye}
\bibfield{author}{\bibinfo{person}{Kajta Hofmann}, \bibinfo{person}{Bhaskar
  Mitra}, \bibinfo{person}{Filip Radlinski}, {and} \bibinfo{person}{Milad
  Shokouhi}.} \bibinfo{year}{2014}\natexlab{}.
\newblock \showarticletitle{An eye-tracking study of user interactions with
  query auto completion}. In \bibinfo{booktitle}{\emph{Proceedings of the 23rd
  ACM International Conference on Conference on Information and Knowledge
  Management}}. \bibinfo{pages}{549--558}.
\newblock


\bibitem[Huang et~al\mbox{.}(2003)]%
        {huang2003relevant}
\bibfield{author}{\bibinfo{person}{Chien-Kang Huang}, \bibinfo{person}{Lee-Feng
  Chien}, {and} \bibinfo{person}{Yen-Jen Oyang}.}
  \bibinfo{year}{2003}\natexlab{}.
\newblock \showarticletitle{Relevant term suggestion in interactive web search
  based on contextual information in query session logs}.
\newblock \bibinfo{journal}{\emph{Journal of the American Society for
  Information Science and Technology}} \bibinfo{volume}{54},
  \bibinfo{number}{7} (\bibinfo{year}{2003}), \bibinfo{pages}{638--649}.
\newblock


\bibitem[Jaech and Ostendorf(2018)]%
        {jaech2018personalized}
\bibfield{author}{\bibinfo{person}{Aaron Jaech} {and} \bibinfo{person}{Mari
  Ostendorf}.} \bibinfo{year}{2018}\natexlab{}.
\newblock \showarticletitle{Personalized language model for query
  auto-completion}.
\newblock \bibinfo{journal}{\emph{arXiv preprint arXiv:1804.09661}}
  (\bibinfo{year}{2018}).
\newblock


\bibitem[Jiang et~al\mbox{.}(2018)]%
        {jiang2018neural}
\bibfield{author}{\bibinfo{person}{Danyang Jiang}, \bibinfo{person}{Wanyu
  Chen}, \bibinfo{person}{Fei Cai}, {and} \bibinfo{person}{Honghui Chen}.}
  \bibinfo{year}{2018}\natexlab{}.
\newblock \showarticletitle{Neural attentive personalization model for query
  auto-completion}. In \bibinfo{booktitle}{\emph{2018 IEEE 3rd Advanced
  Information Technology, Electronic and Automation Control Conference
  (IAEAC)}}. IEEE, \bibinfo{pages}{725--730}.
\newblock


\bibitem[Jiang and Cheng(2016)]%
        {jiang2016classifying}
\bibfield{author}{\bibinfo{person}{Jyun-Yu Jiang} {and} \bibinfo{person}{Pu-Jen
  Cheng}.} \bibinfo{year}{2016}\natexlab{}.
\newblock \showarticletitle{Classifying user search intents for query
  auto-completion}. In \bibinfo{booktitle}{\emph{Proceedings of the 2016 ACM
  International Conference on the Theory of Information Retrieval}}.
  \bibinfo{pages}{49--58}.
\newblock


\bibitem[Jiang et~al\mbox{.}(2014)]%
        {jiang2014learning}
\bibfield{author}{\bibinfo{person}{Jyun-Yu Jiang}, \bibinfo{person}{Yen-Yu Ke},
  \bibinfo{person}{Pao-Yu Chien}, {and} \bibinfo{person}{Pu-Jen Cheng}.}
  \bibinfo{year}{2014}\natexlab{}.
\newblock \showarticletitle{Learning user reformulation behavior for query
  auto-completion}. In \bibinfo{booktitle}{\emph{Proceedings of the 37th
  international ACM SIGIR conference on Research \& development in information
  retrieval}}. \bibinfo{pages}{445--454}.
\newblock


\bibitem[Jiang and Wang(2018)]%
        {jiang2018rin}
\bibfield{author}{\bibinfo{person}{Jyun-Yu Jiang} {and} \bibinfo{person}{Wei
  Wang}.} \bibinfo{year}{2018}\natexlab{}.
\newblock \showarticletitle{RIN: Reformulation inference network for
  context-aware query suggestion}. In \bibinfo{booktitle}{\emph{Proceedings of
  the 27th ACM International Conference on Information and Knowledge
  Management}}. \bibinfo{pages}{197--206}.
\newblock


\bibitem[Ke et~al\mbox{.}(2017)]%
        {ke2017lightgbm}
\bibfield{author}{\bibinfo{person}{Guolin Ke}, \bibinfo{person}{Qi Meng},
  \bibinfo{person}{Thomas Finley}, \bibinfo{person}{Taifeng Wang},
  \bibinfo{person}{Wei Chen}, \bibinfo{person}{Weidong Ma},
  \bibinfo{person}{Qiwei Ye}, {and} \bibinfo{person}{Tie-Yan Liu}.}
  \bibinfo{year}{2017}\natexlab{}.
\newblock \showarticletitle{Lightgbm: A highly efficient gradient boosting
  decision tree}.
\newblock \bibinfo{journal}{\emph{Advances in neural information processing
  systems}}  \bibinfo{volume}{30} (\bibinfo{year}{2017}),
  \bibinfo{pages}{3146--3154}.
\newblock


\bibitem[Kingma and Ba(2014)]%
        {kingma2014adam}
\bibfield{author}{\bibinfo{person}{Diederik~P Kingma} {and}
  \bibinfo{person}{Jimmy Ba}.} \bibinfo{year}{2014}\natexlab{}.
\newblock \showarticletitle{Adam: A method for stochastic optimization}.
\newblock \bibinfo{journal}{\emph{arXiv preprint arXiv:1412.6980}}
  (\bibinfo{year}{2014}).
\newblock


\bibitem[Kulkarni et~al\mbox{.}(2023)]%
        {KulkarniMGF23}
\bibfield{author}{\bibinfo{person}{Hrishikesh Kulkarni}, \bibinfo{person}{Sean
  MacAvaney}, \bibinfo{person}{Nazli Goharian}, {and} \bibinfo{person}{Ophir
  Frieder}.} \bibinfo{year}{2023}\natexlab{}.
\newblock \showarticletitle{Lexically-Accelerated Dense Retrieval}. In
  \bibinfo{booktitle}{\emph{Proceedings of the 46th International {ACM} {SIGIR}
  Conference on Research and Development in Information Retrieval, {SIGIR}
  2023, Taipei, Taiwan, July 23-27, 2023}}. \bibinfo{pages}{152--162}.
\newblock
\urldef\tempurl%
\url{https://doi.org/10.1145/3539618.3591715}
\showURL{%
\tempurl}


\bibitem[Li and Zhang(2023)]%
        {LiZ23}
\bibfield{author}{\bibinfo{person}{Jun Li} {and} \bibinfo{person}{Ge Zhang}.}
  \bibinfo{year}{2023}\natexlab{}.
\newblock \showarticletitle{Fragment and Integrate Network {(FIN):} {A} Novel
  Spatial-Temporal Modeling Based on Long Sequential Behavior for Online Food
  Ordering Click-Through Rate Prediction}. In
  \bibinfo{booktitle}{\emph{Proceedings of the 32nd {ACM} International
  Conference on Information and Knowledge Management, {CIKM} 2023, Birmingham,
  United Kingdom, October 21-25, 2023}}. \bibinfo{pages}{4688--4694}.
\newblock
\urldef\tempurl%
\url{https://doi.org/10.1145/3583780.3615478}
\showDOI{\tempurl}


\bibitem[Liu et~al\mbox{.}(2023)]%
        {00060HSMZ0G23}
\bibfield{author}{\bibinfo{person}{Shuchang Liu}, \bibinfo{person}{Qingpeng
  Cai}, \bibinfo{person}{Zhankui He}, \bibinfo{person}{Bowen Sun},
  \bibinfo{person}{Julian~J. McAuley}, \bibinfo{person}{Dong Zheng},
  \bibinfo{person}{Peng Jiang}, {and} \bibinfo{person}{Kun Gai}.}
  \bibinfo{year}{2023}\natexlab{}.
\newblock \showarticletitle{Generative Flow Network for Listwise
  Recommendation}. In \bibinfo{booktitle}{\emph{Proceedings of the 29th {ACM}
  {SIGKDD} Conference on Knowledge Discovery and Data Mining, {KDD} 2023, Long
  Beach, CA, USA, August 6-10, 2023}}. \bibinfo{pages}{1524--1534}.
\newblock
\urldef\tempurl%
\url{https://doi.org/10.1145/3580305.3599364}
\showDOI{\tempurl}


\bibitem[Mikolov et~al\mbox{.}(2013)]%
        {mikolov2013distributed}
\bibfield{author}{\bibinfo{person}{Tomas Mikolov}, \bibinfo{person}{Ilya
  Sutskever}, \bibinfo{person}{Kai Chen}, \bibinfo{person}{Greg~S Corrado},
  {and} \bibinfo{person}{Jeff Dean}.} \bibinfo{year}{2013}\natexlab{}.
\newblock \showarticletitle{Distributed representations of words and phrases
  and their compositionality}. In \bibinfo{booktitle}{\emph{Advances in neural
  information processing systems}}. \bibinfo{pages}{3111--3119}.
\newblock


\bibitem[Mitra and Craswell(2015)]%
        {mitra2015query}
\bibfield{author}{\bibinfo{person}{Bhaskar Mitra} {and} \bibinfo{person}{Nick
  Craswell}.} \bibinfo{year}{2015}\natexlab{}.
\newblock \showarticletitle{Query auto-completion for rare prefixes}. In
  \bibinfo{booktitle}{\emph{Proceedings of the 24th ACM international on
  conference on information and knowledge management}}.
  \bibinfo{pages}{1755--1758}.
\newblock


\bibitem[Mitra et~al\mbox{.}(2014)]%
        {mitra2014user}
\bibfield{author}{\bibinfo{person}{Bhaskar Mitra}, \bibinfo{person}{Milad
  Shokouhi}, \bibinfo{person}{Filip Radlinski}, {and} \bibinfo{person}{Katja
  Hofmann}.} \bibinfo{year}{2014}\natexlab{}.
\newblock \showarticletitle{On user interactions with query auto-completion}.
  In \bibinfo{booktitle}{\emph{Proceedings of the 37th international ACM SIGIR
  conference on Research \& development in information retrieval}}.
  \bibinfo{pages}{1055--1058}.
\newblock


\bibitem[Park et~al\mbox{.}(2023)]%
        {ParkKCHYCLRYCC23}
\bibfield{author}{\bibinfo{person}{Chung Park}, \bibinfo{person}{Taesan Kim},
  \bibinfo{person}{Taekyoon Choi}, \bibinfo{person}{Junui Hong},
  \bibinfo{person}{Yelim Yu}, \bibinfo{person}{Mincheol Cho},
  \bibinfo{person}{Kyunam Lee}, \bibinfo{person}{Sungil Ryu},
  \bibinfo{person}{Hyungjun Yoon}, \bibinfo{person}{Minsung Choi}, {and}
  \bibinfo{person}{Jaegul Choo}.} \bibinfo{year}{2023}\natexlab{}.
\newblock \showarticletitle{Cracking the Code of Negative Transfer: {A}
  Cooperative Game Theoretic Approach for Cross-Domain Sequential
  Recommendation}. In \bibinfo{booktitle}{\emph{Proceedings of the 32nd {ACM}
  International Conference on Information and Knowledge Management, {CIKM}
  2023, Birmingham, United Kingdom, October 21-25, 2023}}.
  \bibinfo{pages}{2024--2033}.
\newblock
\urldef\tempurl%
\url{https://doi.org/10.1145/3583780.3614828}
\showDOI{\tempurl}


\bibitem[Park and Chiba(2017)]%
        {park2017neural}
\bibfield{author}{\bibinfo{person}{Dae~Hoon Park} {and} \bibinfo{person}{Rikio
  Chiba}.} \bibinfo{year}{2017}\natexlab{}.
\newblock \showarticletitle{A neural language model for query auto-completion}.
  In \bibinfo{booktitle}{\emph{Proceedings of the 40th International ACM SIGIR
  Conference on Research and Development in Information Retrieval}}.
  \bibinfo{pages}{1189--1192}.
\newblock


\bibitem[Pennington et~al\mbox{.}(2014)]%
        {pennington2014glove}
\bibfield{author}{\bibinfo{person}{Jeffrey Pennington},
  \bibinfo{person}{Richard Socher}, {and} \bibinfo{person}{Christopher~D
  Manning}.} \bibinfo{year}{2014}\natexlab{}.
\newblock \showarticletitle{Glove: Global vectors for word representation}. In
  \bibinfo{booktitle}{\emph{Proceedings of the 2014 conference on empirical
  methods in natural language processing (EMNLP)}}.
  \bibinfo{pages}{1532--1543}.
\newblock


\bibitem[Peters et~al\mbox{.}(2018)]%
        {peters2018deep}
\bibfield{author}{\bibinfo{person}{Matthew~E Peters}, \bibinfo{person}{Mark
  Neumann}, \bibinfo{person}{Mohit Iyyer}, \bibinfo{person}{Matt Gardner},
  \bibinfo{person}{Christopher Clark}, \bibinfo{person}{Kenton Lee}, {and}
  \bibinfo{person}{Luke Zettlemoyer}.} \bibinfo{year}{2018}\natexlab{}.
\newblock \showarticletitle{Deep contextualized word representations}.
\newblock \bibinfo{journal}{\emph{arXiv preprint arXiv:1802.05365}}
  (\bibinfo{year}{2018}).
\newblock


\bibitem[Popel and Bojar(2018)]%
        {popel2018training}
\bibfield{author}{\bibinfo{person}{Martin Popel} {and}
  \bibinfo{person}{Ond{\v{r}}ej Bojar}.} \bibinfo{year}{2018}\natexlab{}.
\newblock \showarticletitle{Training tips for the transformer model}.
\newblock \bibinfo{journal}{\emph{arXiv preprint arXiv:1804.00247}}
  (\bibinfo{year}{2018}).
\newblock


\bibitem[Tuan and Phuong(2017)]%
        {tuan20173d}
\bibfield{author}{\bibinfo{person}{Trinh~Xuan Tuan} {and}
  \bibinfo{person}{Tu~Minh Phuong}.} \bibinfo{year}{2017}\natexlab{}.
\newblock \showarticletitle{3D convolutional networks for session-based
  recommendation with content features}. In
  \bibinfo{booktitle}{\emph{Proceedings of the eleventh ACM conference on
  recommender systems}}. \bibinfo{pages}{138--146}.
\newblock


\bibitem[Van~Loan and Golub(1996)]%
        {van1996matrix}
\bibfield{author}{\bibinfo{person}{Charles~F Van~Loan} {and} \bibinfo{person}{G
  Golub}.} \bibinfo{year}{1996}\natexlab{}.
\newblock \showarticletitle{Matrix computations (Johns Hopkins studies in
  mathematical sciences)}.
\newblock  (\bibinfo{year}{1996}).
\newblock


\bibitem[Vaswani et~al\mbox{.}(2017)]%
        {vaswani2017attention}
\bibfield{author}{\bibinfo{person}{Ashish Vaswani}, \bibinfo{person}{Noam
  Shazeer}, \bibinfo{person}{Niki Parmar}, \bibinfo{person}{Jakob Uszkoreit},
  \bibinfo{person}{Llion Jones}, \bibinfo{person}{Aidan~N Gomez},
  \bibinfo{person}{{\L}ukasz Kaiser}, {and} \bibinfo{person}{Illia
  Polosukhin}.} \bibinfo{year}{2017}\natexlab{}.
\newblock \showarticletitle{Attention is all you need}. In
  \bibinfo{booktitle}{\emph{Advances in neural information processing
  systems}}. \bibinfo{pages}{5998--6008}.
\newblock


\bibitem[Voorhees et~al\mbox{.}(1999)]%
        {voorhees1999trec}
\bibfield{author}{\bibinfo{person}{Ellen~M Voorhees} {et~al\mbox{.}}}
  \bibinfo{year}{1999}\natexlab{}.
\newblock \showarticletitle{The TREC-8 question answering track report}. In
  \bibinfo{booktitle}{\emph{Trec}}, Vol.~\bibinfo{volume}{99}. Citeseer,
  \bibinfo{pages}{77--82}.
\newblock


\bibitem[Wang et~al\mbox{.}(2015)]%
        {wang2015learning}
\bibfield{author}{\bibinfo{person}{Pengfei Wang}, \bibinfo{person}{Jiafeng
  Guo}, \bibinfo{person}{Yanyan Lan}, \bibinfo{person}{Jun Xu},
  \bibinfo{person}{Shengxian Wan}, {and} \bibinfo{person}{Xueqi Cheng}.}
  \bibinfo{year}{2015}\natexlab{}.
\newblock \showarticletitle{Learning hierarchical representation model for
  nextbasket recommendation}. In \bibinfo{booktitle}{\emph{Proceedings of the
  38th International ACM SIGIR conference on Research and Development in
  Information Retrieval}}. \bibinfo{pages}{403--412}.
\newblock


\bibitem[Wang et~al\mbox{.}(2020)]%
        {wang2020efficient}
\bibfield{author}{\bibinfo{person}{Sida Wang}, \bibinfo{person}{Weiwei Guo},
  \bibinfo{person}{Huiji Gao}, {and} \bibinfo{person}{Bo Long}.}
  \bibinfo{year}{2020}\natexlab{}.
\newblock \showarticletitle{Efficient Neural Query Auto Completion}. In
  \bibinfo{booktitle}{\emph{Proceedings of the 29th ACM International
  Conference on Information \& Knowledge Management}}.
  \bibinfo{pages}{2797--2804}.
\newblock


\bibitem[Whiting and Jose(2014)]%
        {whiting2014recent}
\bibfield{author}{\bibinfo{person}{Stewart Whiting} {and}
  \bibinfo{person}{Joemon~M Jose}.} \bibinfo{year}{2014}\natexlab{}.
\newblock \showarticletitle{Recent and robust query auto-completion}. In
  \bibinfo{booktitle}{\emph{Proceedings of the 23rd international conference on
  World wide web}}. \bibinfo{pages}{971--982}.
\newblock


\bibitem[Wu et~al\mbox{.}(2018)]%
        {wu2018query}
\bibfield{author}{\bibinfo{person}{Bin Wu}, \bibinfo{person}{Chenyan Xiong},
  \bibinfo{person}{Maosong Sun}, {and} \bibinfo{person}{Zhiyuan Liu}.}
  \bibinfo{year}{2018}\natexlab{}.
\newblock \showarticletitle{Query suggestion with feedback memory network}. In
  \bibinfo{booktitle}{\emph{Proceedings of the 2018 World Wide Web
  Conference}}. \bibinfo{pages}{1563--1571}.
\newblock


\bibitem[Wu et~al\mbox{.}(2020)]%
        {wu2020sse}
\bibfield{author}{\bibinfo{person}{Liwei Wu}, \bibinfo{person}{Shuqing Li},
  \bibinfo{person}{Cho-Jui Hsieh}, {and} \bibinfo{person}{James Sharpnack}.}
  \bibinfo{year}{2020}\natexlab{}.
\newblock \showarticletitle{SSE-PT: Sequential recommendation via personalized
  transformer}. In \bibinfo{booktitle}{\emph{Fourteenth ACM Conference on
  Recommender Systems}}. \bibinfo{pages}{328--337}.
\newblock


\bibitem[Yang et~al\mbox{.}(2023)]%
        {YangHXH23}
\bibfield{author}{\bibinfo{person}{Yuhao Yang}, \bibinfo{person}{Chao Huang},
  \bibinfo{person}{Lianghao Xia}, {and} \bibinfo{person}{Chunzhen Huang}.}
  \bibinfo{year}{2023}\natexlab{}.
\newblock \showarticletitle{Knowledge Graph Self-Supervised Rationalization for
  Recommendation}. In \bibinfo{booktitle}{\emph{Proceedings of the 29th {ACM}
  {SIGKDD} Conference on Knowledge Discovery and Data Mining, {KDD} 2023, Long
  Beach, CA, USA, August 6-10, 2023}}. \bibinfo{pages}{3046--3056}.
\newblock
\urldef\tempurl%
\url{https://doi.org/10.1145/3580305.3599400}
\showDOI{\tempurl}


\bibitem[Yin et~al\mbox{.}(2020)]%
        {yin2020learning}
\bibfield{author}{\bibinfo{person}{Di Yin}, \bibinfo{person}{Jiwei Tan},
  \bibinfo{person}{Zhe Zhang}, \bibinfo{person}{Hongbo Deng},
  \bibinfo{person}{Shujian Huang}, {and} \bibinfo{person}{Jiajun Chen}.}
  \bibinfo{year}{2020}\natexlab{}.
\newblock \showarticletitle{Learning to Generate Personalized Query
  Auto-Completions via a Multi-View Multi-Task Attentive Approach}. In
  \bibinfo{booktitle}{\emph{Proceedings of the 26th ACM SIGKDD International
  Conference on Knowledge Discovery \& Data Mining}}.
  \bibinfo{pages}{2998--3007}.
\newblock


\bibitem[Zamani and Croft(2016)]%
        {zamani2016estimating}
\bibfield{author}{\bibinfo{person}{Hamed Zamani} {and} \bibinfo{person}{W~Bruce
  Croft}.} \bibinfo{year}{2016}\natexlab{}.
\newblock \showarticletitle{Estimating embedding vectors for queries}. In
  \bibinfo{booktitle}{\emph{Proceedings of the 2016 ACM International
  Conference on the Theory of Information Retrieval}}.
  \bibinfo{pages}{123--132}.
\newblock


\bibitem[Zamani and Croft(2017)]%
        {zamani2017relevance}
\bibfield{author}{\bibinfo{person}{Hamed Zamani} {and} \bibinfo{person}{W~Bruce
  Croft}.} \bibinfo{year}{2017}\natexlab{}.
\newblock \showarticletitle{Relevance-based word embedding}. In
  \bibinfo{booktitle}{\emph{Proceedings of the 40th International ACM SIGIR
  Conference on Research and Development in Information Retrieval}}.
  \bibinfo{pages}{505--514}.
\newblock


\bibitem[Zhang et~al\mbox{.}(2018)]%
        {zhang2018next}
\bibfield{author}{\bibinfo{person}{Shuai Zhang}, \bibinfo{person}{Yi Tay},
  \bibinfo{person}{Lina Yao}, {and} \bibinfo{person}{Aixin Sun}.}
  \bibinfo{year}{2018}\natexlab{}.
\newblock \showarticletitle{Next item recommendation with self-attention}.
\newblock \bibinfo{journal}{\emph{arXiv preprint arXiv:1808.06414}}
  (\bibinfo{year}{2018}).
\newblock


\bibitem[Zhang et~al\mbox{.}(2023a)]%
        {ZhangSFWW0023}
\bibfield{author}{\bibinfo{person}{Yang Zhang}, \bibinfo{person}{Tianhao Shi},
  \bibinfo{person}{Fuli Feng}, \bibinfo{person}{Wenjie Wang},
  \bibinfo{person}{Dingxian Wang}, \bibinfo{person}{Xiangnan He}, {and}
  \bibinfo{person}{Yongdong Zhang}.} \bibinfo{year}{2023}\natexlab{a}.
\newblock \showarticletitle{Reformulating {CTR} Prediction: Learning Invariant
  Feature Interactions for Recommendation}. In
  \bibinfo{booktitle}{\emph{Proceedings of the 46th International {ACM} {SIGIR}
  Conference on Research and Development in Information Retrieval, {SIGIR}
  2023, Taipei, Taiwan, July 23-27, 2023}}. \bibinfo{pages}{1386--1395}.
\newblock
\urldef\tempurl%
\url{https://doi.org/10.1145/3539618.3591755}
\showURL{%
\tempurl}


\bibitem[Zhang et~al\mbox{.}(2023b)]%
        {ZhangWLZQHJPLS23}
\bibfield{author}{\bibinfo{person}{Yinan Zhang}, \bibinfo{person}{Pei Wang},
  \bibinfo{person}{Congcong Liu}, \bibinfo{person}{Xiwei Zhao},
  \bibinfo{person}{Hao Qi}, \bibinfo{person}{Jie He}, \bibinfo{person}{Junsheng
  Jin}, \bibinfo{person}{Changping Peng}, \bibinfo{person}{Zhangang Lin}, {and}
  \bibinfo{person}{Jingping Shao}.} \bibinfo{year}{2023}\natexlab{b}.
\newblock \showarticletitle{{BI-GCN:} Bilateral Interactive Graph Convolutional
  Network for Recommendation}. In \bibinfo{booktitle}{\emph{Proceedings of the
  32nd {ACM} International Conference on Information and Knowledge Management,
  {CIKM} 2023, Birmingham, United Kingdom, October 21-25, 2023}}.
  \bibinfo{pages}{4410--4414}.
\newblock
\urldef\tempurl%
\url{https://doi.org/10.1145/3583780.3615232}
\showDOI{\tempurl}


\bibitem[Zhao et~al\mbox{.}(2018)]%
        {zhao2018generalizing}
\bibfield{author}{\bibinfo{person}{Jinman Zhao}, \bibinfo{person}{Sidharth
  Mudgal}, {and} \bibinfo{person}{Yingyu Liang}.}
  \bibinfo{year}{2018}\natexlab{}.
\newblock \showarticletitle{Generalizing word embeddings using bag of
  subwords}.
\newblock \bibinfo{journal}{\emph{arXiv preprint arXiv:1809.04259}}
  (\bibinfo{year}{2018}).
\newblock


\bibitem[Zhong et~al\mbox{.}(2020)]%
        {zhong2020personalized}
\bibfield{author}{\bibinfo{person}{Jianling Zhong}, \bibinfo{person}{Weiwei
  Guo}, \bibinfo{person}{Huiji Gao}, {and} \bibinfo{person}{Bo Long}.}
  \bibinfo{year}{2020}\natexlab{}.
\newblock \showarticletitle{Personalized Query Suggestions}. In
  \bibinfo{booktitle}{\emph{Proceedings of the 43rd International ACM SIGIR
  Conference on Research and Development in Information Retrieval}}.
  \bibinfo{pages}{1645--1648}.
\newblock


\bibitem[Zhou et~al\mbox{.}(2018)]%
        {zhou2018deep}
\bibfield{author}{\bibinfo{person}{Guorui Zhou}, \bibinfo{person}{Xiaoqiang
  Zhu}, \bibinfo{person}{Chenru Song}, \bibinfo{person}{Ying Fan},
  \bibinfo{person}{Han Zhu}, \bibinfo{person}{Xiao Ma},
  \bibinfo{person}{Yanghui Yan}, \bibinfo{person}{Junqi Jin},
  \bibinfo{person}{Han Li}, {and} \bibinfo{person}{Kun Gai}.}
  \bibinfo{year}{2018}\natexlab{}.
\newblock \showarticletitle{Deep interest network for click-through rate
  prediction}. In \bibinfo{booktitle}{\emph{Proceedings of the 24th ACM SIGKDD
  International Conference on Knowledge Discovery \& Data Mining}}.
  \bibinfo{pages}{1059--1068}.
\newblock


\bibitem[Zhu et~al\mbox{.}(2019)]%
        {zhu2019systematic}
\bibfield{author}{\bibinfo{person}{Yi Zhu}, \bibinfo{person}{Ivan Vuli{\'c}},
  {and} \bibinfo{person}{Anna Korhonen}.} \bibinfo{year}{2019}\natexlab{}.
\newblock \showarticletitle{A systematic study of leveraging subword
  information for learning word representations}.
\newblock \bibinfo{journal}{\emph{arXiv preprint arXiv:1904.07994}}
  (\bibinfo{year}{2019}).
\newblock


\end{thebibliography}

\clearpage\end{CJK*}
\end{document}